\documentclass{elsart}
\usepackage{epsfig}
\usepackage{amssymb}

\begin{document}
\begin{frontmatter}

\title{Solitons in Bose-Einstein condensates trapped in a double-well potential}
\author{Valery S. Shchesnovich\corauthref{cor}},
\corauth[cor]{Corresponding author.} \ead{valery@ift.unesp.br}
\author{Boris A. Malomed\thanksref{leave}}, and
\thanks[leave]{Permanent address: Department of Interdisciplinary Studies,
Faculty of Engineering, Tel Aviv University, Tel Aviv 69978, Israel}
\author{Roberto A. Kraenkel}
\address{Instituto de F{i}sica Te\'{o}rica, Universidade Estadual Paulista, Rua
Pamplona 145, 01405-900 S\~{a}o Paulo, Brazil}

\begin{abstract}
We investigate, analytically and numerically, families of bright solitons in a
system of two linearly coupled nonlinear Schr\"{o}dinger/Gross-Pitaevskii
equations, describing two Bose-Einstein condensates trapped in an asymmetric
double-well potential, in particular, when the scattering lengths in the
condensates have arbitrary magnitudes and opposite signs. The solitons are
found to exist everywhere where they are permitted by the dispersion law. Using
the Vakhitov-Kolokolov criterion and numerical methods, we show that, except
for small regions in the parameter space, the solitons are stable to small
perturbations. Some of them feature self-trapping of almost all the atoms in
the condensate with no atomic interaction or weak repulsion coupled to the
self-attractive condensate. An unusual bifurcation is found, when the soliton
bifurcates from the zero solution without a visible jump in the shape, but with
a jump in the number of trapped atoms.  By means of numerical simulations, it
is found that, depending on values of the parameters and the initial
perturbation, unstable solitons either give rise to breathers or completely
break down into incoherent waves (``radiation"). A version of the model with
the self-attraction in both components, which applies to the description of
dual-core fibers in nonlinear optics, is considered too, and new results are
obtained for this much studied system.
\end{abstract}

\begin{keyword}
{solitons in Bose-Einstein condensates, coupled Nonlinear Schr\"odinger
equations, soliton stability} \PACS{05.45.Yv,03.75.Fi,42.81.Dp}
\end{keyword}

\end{frontmatter}

\section{Introduction}
\label{intro}

Experimental observation of Bose-Einstein condensates (BECs) in trapped dilute
gases \cite{exp1,exp2,exp3,exp4} had opened new exciting possibilities for
manifestations of nonlinear phenomena in various geometries. Indeed, in the
mean-field approximation (which usually applies to a great accuracy), the order
parameter, which can be identified with the single-atom wave function, obeys
the nonlinear Schr\"{o}dinger (NLS) equation with an external potential, also
known as the Gross-Pitaevskii (GP) equation \cite{GPE}. By varying the trap
potential, the shape of the condensate can be tailored to sundry geometries, in
particular, the shape of the condensate may be approximately circular, i.e.,
the condensate itself may be effectively two-dimensional (2D) or
one-dimensional (1D, alias cigar-shaped) \cite{rev1}.

It was recently shown that not only the geometry of the condensate, but also
the magnitude and the sign of the scattering length, which determines
interactions between atoms in the condensate and, thus, the nonlinear term in
the corresponding GP equation, can be manipulated by varying the external
magnetic field near the Feshbach resonance \cite{fesh}. This opens additional
possibilities to control the quantum macroscopic dynamics of BECs.

The NLS equation is a fundamental model for many physical media. A well-known
application of the 1D NLS equation is the description of the pulse propagation
in nonlinear optical fibers \cite{optfiber}. Similar to optics, where bright
and dark solitons are supported by the focusing and defocusing nonlinearity,
respectively, in BECs the $s$-wave scattering interaction between atoms is a
similar determining factor. Therefore, condensates constrained to the
one-dimensional shape can form dark or bright solitons. Dark solitons were
found in condensates with repulsive interactions
\cite{dark1,dark2,dark3,dark4}, while condensates with attractive interaction
were recently experimentally shown to form stable bright solitons
\cite{bright1}. The appearance of solitons is one of the most interesting
manifestations of nonlinear dynamics, and in the case of BECs it is of
paramount interest, as in this case the solitons represent self-localized
``waves of matter''.

The similarity of the NLS and GP equations suggests that many nonlinear
phenomena in BECs may have their counterparts in nonlinear optics, in
particular, in fibers. However, due to unique manageability of the properties
of BECs, new setups can be studied, which were not realized in nonlinear
optics. One of such setups is related to the above-mentioned possibility of the
effective control of the scattering length in BECs, i.e., the sign of the
nonlinearity in the governing equations.

Anticipating such experiments, in the present paper we find and study,
analytically and numerically, a new family of solitons in two weakly coupled
effectively 1D condensates, trapped in a double-well magnetic potential, when
the $s$-scattering lengths of the two condensates have different magnitudes,
with particular emphasis on the case when the scattering lengths have opposite
signs.

The spatially separated condensates can be created by focusing a
far-off-resonant intense laser beam, which generates a repulsive optical dipole
force, into the center of a magnetic trap. We assume that the condensates have
the cigar-like shapes, with the transversal  dimensions being strongly
constrained by the trap, see Fig.~1. The chemical-potential difference between
the traps, $\mu _{0}$, can be managed, moving the position of the
barrier-generating laser beam by means of electro-optic or acousto-optic
modulators (see also Ref.~\cite{modul}). Such a setup was realized in
experiments (see, for instance, Ref. \cite{exp4}). We neglect a variation of
the trap potential along the longitudinal (say, $x$) direction assuming it to
be weak, so that for localized solutions well inside the trap the external
potential can be considered as flat. Finally, as we are interested in the
soliton solutions, we take into account the kinetic energy contribution.

\begin{figure}[f]
\psfig{file=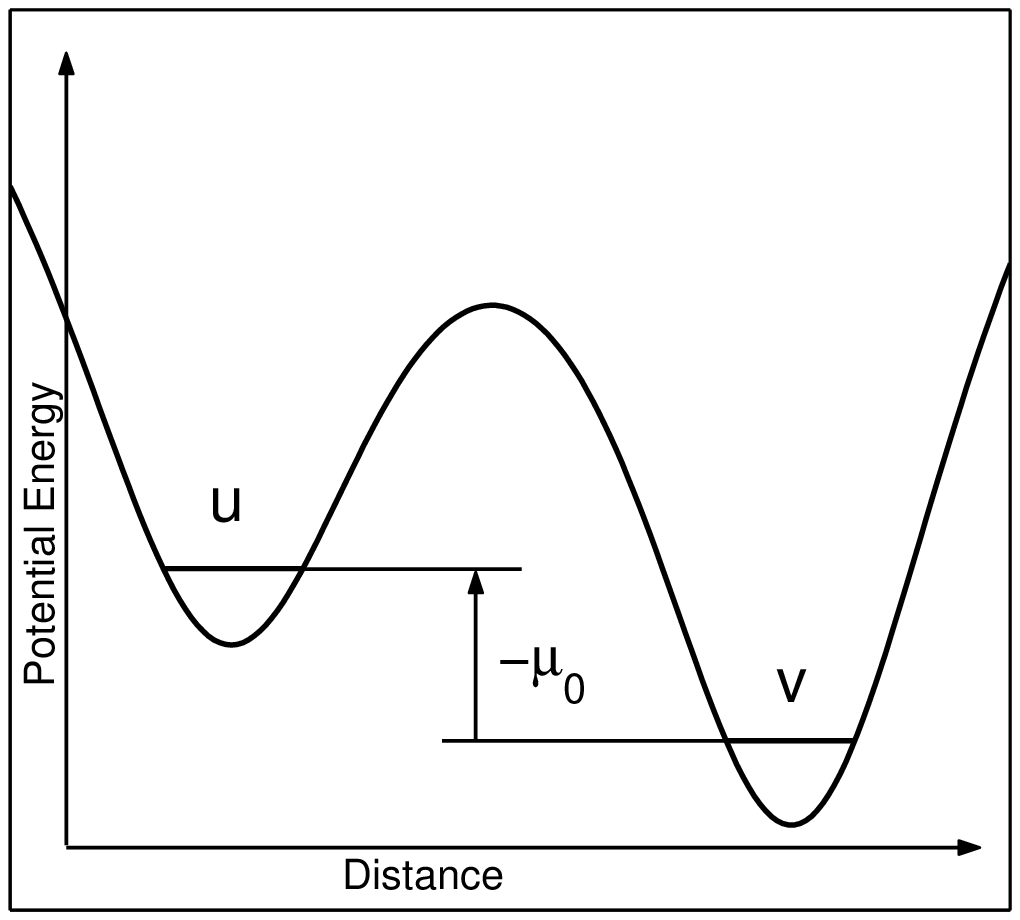,height=0.3\textwidth,width=0.3\textwidth} \caption{A
scematic transverse shape of the trapping potential. The two condensates are
designated by the symbols $u$ and $v$. The difference of the chemical
potentials between them is $\protect\mu_0$.}
\end{figure}

The corresponding coupled-mode equations, after obvious scaling
transformations, may be cast in the following dimensionless form
\begin{eqnarray}
&&i\partial _{t}u+\partial _{x}^{2}u+|u|^{2}u+v=0,  \label{E1} \\
&&i\partial _{t}v+\partial _{x}^{2}v-(\mu _{0}+a|v|^{2})v+u=0,  \label{E2}
\end{eqnarray}
where $\mu _{0}$ accounts for a difference of the chemical potentials between
the two condensates, described by the order parameters $u$ and $v$. It is
assumed that the nonlinear interaction is always attractive in the
$u$-condensate, while in the $v$-condensate the strength and sign of the
nonlinearity are controlled by the coefficient $a$ (e.g., repulsive
nonlinearity if $a>0$). The derivation of the couple-mode equations similar to
Eqs.~(\ref{E1})-(\ref{E2}) from the corresponding GP equation was discussed in
many works (consult, for instance, Refs.~\cite{tunnellimit,tunnel1,tunnel2}),
the outline is placed in appendix~\ref{deriv}. Thus, the system
(\ref{E1})-(\ref{E2}) realizes an interesting interplay between the dispersion
(i.e., the kinetic energy contribution), self-focusing and self-defocusing
nonlinearities, linear coupling and the potential shift, which deserves the
study.

Dynamics of two BECs in a magnetic trap was studied before. For instance, existence
of a macroscopic quantum-phase difference, experimentally demonstrated in
Ref.~\cite{exp4}, was used to study the coherent atomic tunnelling between two
weakly coupled BECs confined in a double-well potential \cite{tunnel1,tunnel2}. In
Ref.~\cite{tunnellimit} it was shown that the coherent oscillations due to
tunnelling are suppressed when the number of atoms exceeds a critical value. The
effect of the trap oscillations on the atomic tunnelling between two BECs was
analyzed in Ref.~\cite{timedep}. In Ref.~\cite{modopt} an analogy with the
nonlinear-optical directional fiber couplers  was used to account for the kinetic
terms in governing coupled NLS/GP equations (for a review of nonlinear dynamics in
optical fiber couplers see section~6 of Ref.~\cite{NewReview}; this analogy will
play an important in the present work too). Further use of the analogy with
guided-wave optics led to the development of a nonlinear collective-mode theory for
BECs trapped in an external potential \cite{couplmod}. In Ref.~\cite{nonpert},
nonlinear modes in the form of chains of bright and dark solitons, which have no
linear counterparts, were shown to be stationary solutions to the NLS equation with
a multi-well external potential. A related model, based on a multi-component (i.e.,
vector) NLS equation, that describes repulsively interacting two-component BECs,
possesses a coupled dark-bright soliton solution \cite{darkbright}.

For $a<0$ (i.e., when the nonlinearities are focusing in both subsystems)
equations similar to the system (\ref{E1})-(\ref{E2}) have been studied in
connection with the nonlinear dynamics in the dual-core optical fibers (see the
above-mentioned review \cite{NewReview} and original
papers~\cite{A1,A2,A3,M1,K1,K2}). However, the system (\ref{E1})-(\ref{E2})
with $a\geq 0$, i.e., with opposite signs of the nonlinearity in the two
condensates, was not considered before. Besides its relevance to the
description of the coupled BECs, as argued above, the study of the model with
positive $a$ is of general interest: as we demonstrate in this work, it gives
rise to interesting soliton solutions which exhibit some unusual properties
(see figures~4 and~6 below, for instance). Whereas letting the parameter $a$ to
have {\emph arbitrary} negative values we also uncover novel features of the
solitons and bifurcations  studied before for the case of $a=-1$.

The paper is organized as follows. In the next section, we aim to identify a
domain of the soliton existence (mainly for the case of $a\geq 0$) by deriving
equations and inequalities which must be satisfied by the soliton solutions. In
this section, we also find particular (sech-type) soliton solutions for $a<0$.
Generic numerically found soliton solutions are presented in
section~\ref{numerics}. In particular, a noteworthy result reported in this
section is the occurrence of an unusual bifurcation (which was found for
$a=1$): a soliton may have its amplitude vanishing and width simultaneously
diverging, while the number of atoms in this configuration remains finite. In
section~\ref{stab}, we derive a criterion of the Vakhitov-Kolokolov (VK) type
for the soliton stability, and study the fate of unstable solitons subject to
small perturbations by means of direct numerical simulations. As the result, it
is found that, depending on the values of the system parameters and on the form
of the initial small perturbation, the unstable soliton either  gives rise to a
breather (with persistent intrinsic vibrations whose amplitude is dependent on
the slope of the number of trapped atoms vs. the chemical potential) or
completely decays into radiation. The concluding section summarizes results
obtained in the work.

\section{The region of soliton existence and analytical
solutions}

\label{sec1}

To search for the stationary solitary-pulse solutions to Eqs. (\ref{E1}) and
(\ref{E2}) we set
\begin{equation}
u(t,x)=e^{-i\mu t}U(x),\quad v(t,x)=e^{-i\mu t}V(x),  \label{UV}
\end{equation}
where $\mu $ is the normalized (dimensionless) chemical potential, and $U(x)$
and $V(x)$ are real functions vanishing as $x\rightarrow \pm \infty $. Thus we
arrive at a system of two real ordinary differential equations for the pulse
profiles:
\begin{eqnarray}
&&\frac{\mathrm{d}^{2}U}{\mathrm{d}x^{2}}+(\mu +U^{2})U+V=0,
\label{E4U} \\
&&\frac{\mathrm{d}^{2}V}{\mathrm{d}x^{2}}+(\mu -\mu _{0}-aV^{2})V+U=0.
\label{E4V}
\end{eqnarray}
Equations (\ref{E4U})-(\ref{E4V}) were numerically solved, to look for solitons
in a wide domain in the parameter space ($a$, $\mu _{0}$, $\mu $). However,
before proceeding to the numerical solution, some results can be obtained in an
analytical form, which will provide for some insight into the existence of
solitons. The analytical results will make it possible to narrow a domain in
the parameter space where it makes sense to look for solitons numerically.
Moreover, the analytical results will be used to check the numerical solutions.

First of all, considered as a dynamical system, equations (\ref{E4U})-(\ref
{E4V}) possess a Hamiltonian,
\begin{equation}
\mathcal{H}=\left( \frac{\mathrm{d}U}{\mathrm{d}x}\right) ^{2}+\left( \frac{
\mathrm{d}V}{\mathrm{d}x}\right) ^{2}+\left( \mu +\frac{U^{2}}{2}\right)
U^{2}+\left( \mu -\mu _{0}-\frac{aV^{2}}{2}\right) V^{2}+2UV.
\label{Ham}
\end{equation}
Evidently, solutions vanishing as $|x|\rightarrow \infty$ correspond to
$\mathcal{H}=0$. We specify the class of the soliton solutions we look for as
even solutions, $U(-x)=U(x)$, $V(-x)=V(x)$, with a single maximum at $x=0$ (we
aim to consider single-humped, i.e., fundamental solitons). For the
definiteness' sake, we set $U(x=0)\equiv U_{0}>0 $. As the first derivatives of
the fields vanish at the central point, we can derive the following quartic
equation for the soliton amplitudes $U_{0}$ and $V_{0}\equiv V(0)$ from Eqs.
(\ref{E4U}) and (\ref{E4V}):
\begin{equation}
\left( \mu +\frac{U_{0}^{2}}{2}\right) U_{0}^{2}+\left( \mu -\mu _{0}-\frac{
aV_{0}^{2}}{2}\right) V_{0}^{2}+2U_{0}V_{0}=0.  \label{cond1}
\end{equation}

Further, multiplication of Eq.~(\ref{E4U}) by ${\mathrm{d}U}/{\mathrm{d}x}$,
and (\ref{E4V}) by ${\mathrm{d}V}/{\mathrm{d}x}$, and the integration from
$x=0$ to $x=\infty $ leads to the following identities:
\begin{equation}
\left( \mu +\frac{U_{0}^{2}}{2}\right) U_{0}^{2}=2\int\limits_{0}^{\infty }
\mathrm{d}x\,V\frac{\mathrm{d}U}{\mathrm{d}x},\quad \left( \mu -\mu _{0}-\frac{
aV_{0}^{2}}{2}\right) V_{0}^{2}=2\int\limits_{0}^{\infty }\mathrm{d}x\,U\frac{
\mathrm{d}V}{\mathrm{d}x}.  \label{ints}
\end{equation}
In principle, both in-phase ($V_{0}>0$) as out-of-phase ($V_{0}<0$) fundamental
solitons are possible (recall we have set $U_{0}>0$). Evidently, the right-hand
sides of Eqs.~(\ref{ints}) are negative for the in-phase, and positive for the
out-of-phase solitons. Thus, we conclude that the fundamental solitons must
obey two inequalities:
\begin{eqnarray}
U_{0}^{2} &<&-2\mu ,\quad \mathrm{if}\quad V_{0}>0,  \label{inph} \\
aV_{0}^{2} &<&2(\mu -\mu _{0}),\quad \mathrm{if}\quad V_{0}<0.  \label{outph}
\end{eqnarray}
For $a>0$ (the opposite signs of the nonlinearities in the two subsystems),
from these inequalities it follows that, for the in-phase fundamental solitons,
the chemical potential of the $U$-condensate is negative ($\mu <0$), while, for
the out-of-phase solitons, the chemical potential of the $V$-condensate is
positive ($\mu -\mu _{0}>0$).

The domain where the soliton solutions are permitted is determined by the
dispersion law. Assuming the exponential decay $U\sim e^{-kx}$, $V\sim e^{-kx}$
for $\left| x\right| \rightarrow \infty $, and linearizing equations
(\ref{E4U})-(\ref{E4V}), we find two branches of the dispersion relation for
the solitons:
\begin{equation}
k_{1}^{2}=-\mu +\frac{\mu _{0}}{2}-\sqrt{\left( \frac{\mu _{0}}{2}\right)
^{2}+1},\quad k_{2}^{2}=-\mu +\frac{\mu _{0}}{2}+\sqrt{\left( \frac{\mu _{0}
}{2}\right) ^{2}+1}.  \label{disp}
\end{equation}
It is seen that the condition $k_{1}^{2}>0$ implies $\mu <\mu _{0}$, which
excludes the out-of-phase solitons for this branch for positive $a$. The
appealing conclusion that the $k_{1}$-branch corresponds to the in-phase
soliton solutions is confirmed by our numerical results (for both positive and
negative $a$), see the next section. This fact is also used in the proof of the
Vakhitov-Kolokolov stability criterion for the in-phase solitons in
section~\ref{stab}.

Finally, analysis of solutions of Eq. (\ref{cond1}) by means of the
inequalities (\ref{inph}) and (\ref{outph}) shows that, for fixed $a>0$,
$\mu_{0}$, $\mu $, and $U_{0}$, there is one negative [satisfying the condition
(\ref{outph})] and, at most, two positive [satisfying the condition
(\ref{inph})] real solutions for $V_{0}$. Thus, there may be at most two
branches of the in-phase solitons (however, we were able to find, in a
numerical form, only one in-phase soliton solution corresponding to a given
$U_{0}$ for fixed $a$, $\mu _{0}$, and $\mu $, see the next section).

For $a<0$, when the system (\ref{E4U})-(\ref{E4V}) describes two coupled
condensates with attractive interactions, it also provides for a
straightforward generalization of the model describing the (mismatched)
nonlinear dual-core optical fiber. In this context, the soliton solutions were
studied in detail for the case $a=-1$, i.e., equal Kerr coefficients in both
cores \cite{A1,A2,A3,M1,K1,K2}. In this case, the parameter $\mu _{0}$ accounts
for the phase-velocity difference between the cores~\cite{M1,K1}.

For negative $a$, Eqs. (\ref{E4U})-(\ref{E4V}) have special exact soliton
solutions, both in- and out-of-phase ones, which generalize, respectively, the
well-known symmetric and antisymmetric solitons in the standard model of the
dual-core optical fiber \cite{A1}. The exact solutions for the in-phase
solitons are
\begin{equation}
\left(\begin{array}{c}U\\ V\end{array}\right) = \left(\begin{array}{c}A_{1}\\
A_{1}/\sqrt{-a}\end{array}\right)\mathrm{sech}\left(
\frac{A_{1}x}{\sqrt{2}}\right), \quad A_{1}= \sqrt{2}\left( -\mu
-\frac{1}{\sqrt{-a}}\right) ^{1/2},
\label{exactinph}
\end{equation}
which exist in the special case, when $\mu _{0}$ and $a$  are not independent
parameters, but are related as follows:
\begin{equation}
\mu _{0}=\sqrt{-a}-\frac{1}{\sqrt{-a}}\equiv \mu _{0}\left( a\right) ,\quad \mu
<\mu _{\max }^{(1)}\equiv -\frac{1}{\sqrt{-a}}.  \label{max1}
\end{equation}
The out-of-phase solitons are
\begin{equation}
\left(\begin{array}{c}U\\ V\end{array}\right) = \left(\begin{array}{c}A_2\\
-A_2/\sqrt{-a}\end{array}\right)\mathrm{sech}\left(
\frac{A_2x}{\sqrt{2}}\right) ,\quad A_{2}= \sqrt{2}\left( -\mu
+\frac{1}{\sqrt{-a}}\right) ^{1/2},  \label{exactoutph}
\end{equation}
and they exist in the case
\begin{equation}
\mu _{0}=-\mu _{0}\left( a\right) ,\quad \mu <\mu _{\max }^{(2)}\equiv
\frac{1}{ \sqrt{-a}}  \label{max2}
\end{equation}
[recall $\mu _{0}\left( a\right)$ is defined in Eq. (\ref {max1})]. Note that
the in-phase sech-type solitons correspond to the first branch of the
dispersion law (\ref{disp}), $A_{1}^{2}/2=k_{1}^{2}$, while the out-of-phase
solitons correspond to the second branch, $A_{2}^{2}/2=k_{2}^{2} $.
Unfortunately, the solutions (\ref{exactinph}) and (\ref{exactoutph}) cannot be
continued to positive values of $a$.

Some additional exact results can be obtained concerning families of soliton
solutions to Eqs. (\ref{E4U})-(\ref{E4V}) with $a<0$. For instance, by using
the standard bifurcation analysis it is easy to show that there is another
family of the in-phase solitons, bifurcating from the solutions
(\ref{exactinph}) at the point
\begin{equation}
\mu =\mu _{\mathrm{bif}}\equiv -\frac{4-a}{3\sqrt{-a}}\,,  \label{bif}
\end{equation}
if $\mu _{0}$ and $a$ are related as in equation (\ref{max1}) (this bifurcation
is considered in detail numerically in section~\ref{a<0} and  analytically in
section~\ref{stab}). Setting $a=-1$, one recovers the bifurcation which was
found, in terms of the dual-core fiber, in Ref.~\cite{A1} (in this case $\mu
_{0}=0$). Note that due to the relation $\mu _{\mathrm{bif}}<\mu _{\max
}^{(1)}$ (which can be easily checked to hold) this bifurcation is present for
any $a<0$. Actually, the bifurcating in-phase solitons generalize the so-called
$A$-type solitons (defined in Ref.~\cite{A1} for $a=-1$, see also
Refs.~\cite{M1,K1}) for arbitrary (negative) values of $a$.

Another family of soliton solutions, the so-called $B$-type solitons,
bifurcates from the out-of-phase solitons in a non-perturbative way \cite{A1}.
However, the $B$-type solitons were shown to be unstable in the whole domain of
their existence, while the out-of-phase solitons could be stable only in a very
narrow interval \cite {A2}. For this reason, we discard the out-of-phase
solitons from further consideration, so that all the solitons considered below
are implied to be of the in-phase type.

\section{Soliton solutions (numerical analysis)}

\label{numerics}

We will discuss only the normalized quantities corresponding to the system
(\ref{E1})-(\ref{E2}). The numbers of the trapped atoms are also normalized
accordingly, so that our discussion below is general instead of being tied to a
particular shape and size of the external trap.  The actual numbers of trapped
atoms depend heavily on the particulars of the trap according to formula
(\ref{Nphys}) given in appendix~\ref{deriv}. The transformation coefficient between
the physical and ``normalized'' numbers of atoms is estimated there to be on the
order of $10^3$ for the current experimental setups. Thus the soliton solutions we
discuss below correspond to the interval from thousands to a hundred thousands of
the trapped atoms. Those are the characteristic numbers of atoms in the current
experiments. On the other hand, it is not surprising to come up with precisely this
interval in our model, since it is derived under the assumption of the weak
nonlinearity, in which case the upper bound on the number of trapped atoms is
determined by the external trap. {\it To avoid confusion, we note here that below
by the ``number of trapped atoms'' we mean only the number of particles in the
model (\ref{E1})-(\ref{E2})}.

To solve Eqs. (\ref{E4U})-(\ref{E4V}) numerically the following iterative
scheme was adopted
\begin{eqnarray}
&&\frac{\mathrm{d}^{2}U^{(n)}}{\mathrm{d}x^{2}}+\left( U^{(n-1)}\right)
^{2}U^{(n)}+V^{(n)}=-\mu ^{(n)}U^{(n)},  \label{4Un} \\
&&\frac{\mathrm{d}^{2}V^{(n)}}{\mathrm{d}x^{2}}-\left[ \mu _{0}+a\left(
V^{(n-1)}\right) ^{2}\right] V^{(n)}+U^{(n)}=-\mu ^{(n)}V^{(n)},  \label{4Vn}
\end{eqnarray}
i.e. we solve the eigenvalue problem for $\mu ^{(n)}$ at the $n$-th iterative
step. Selecting the lowest eigenvalue at each iterative step leads to
convergence to a soliton solution\footnote{In some cases for $a<0$, when there
are several soliton solutions corresponding to the same  $a$, $\mu_0$, and
$\mu$, we used the shooting method to obtain the branches of solitons, to which
the iterative scheme (\ref{4Un})-(\ref{4Vn}) has poor convergence.} (the use of
this scheme, and the selection of the eigenvalue, are prompted by the
well-known facts from the one-dimensional eigenvalue theory). We used the
Fourier spectral (collocation) method with up to 256 grid points (see Refs.
\cite {Forn,Boyd,Tref} for introduction into the spectral methods). Geometric
convergence of the numerical solution to a soliton was noted for quite
arbitrary initial profiles. As our method required only one of the two
amplitudes to be specified (by appropriate normalization of the eigenfunction
at each step), while the other was the result of the computation, we used the
exact relations (\ref{cond1}) to check the correctness of the numerical
solutions. We also tested our approach, using the explicit soliton solutions
(\ref{exactinph}).

\subsection{Solitons for $a\geq 0$}

In the model with the opposite signs of the nonlinearity in the two
condensates, $a>0$, we have found a family of soliton solutions existing for
\emph{all} values of $a$ and $\mu _{0}$, and \emph{all} the values of the
chemical potential $\mu $ permissible by the dispersion law, i.e., which
satisfy the condition
\begin{equation}
\mu <\mu _{\mathrm{\max }}\equiv \frac{\mu _{0}}{2}-\sqrt{\left( \frac{\mu
_{0}}{2}\right) ^{2}+1}.  \label{rangemu}
\end{equation}
It should be stressed that, in comparison with the previous works done in the
context of nonlinear optics, these are essentially novel solitons, as all the
previously studied cases \cite{A1,A2,A3,M1,K1,K2} assumed the same sign of the
nonlinearity in both cores of the system (although a possibility of the
existence of bright gap solitons was shown in the case when the signs were
opposite in front of the second derivatives \cite{K2}; in optics, this case is
quite possible, corresponding to opposite signs of the group-velocity
dispersion in a dual-core fiber with asymmetric cores, while in the case of BEC
this case makes no sense).

\medskip \noindent \textsl{Case I. $a=0$}

We start with the special case $a=0$, when the second condensate is characterized
by the zero scattering length (no nonlinearity). In figure~2 we plot the total
number of atoms, and the numbers of atoms in each condensate as functions of the
chemical potential $\mu $ for several values of the chemical-potential difference
$\mu _{0}$.  These dependencies, besides being important to quantity the BECs,
determine the stability of the soliton solution: for fixed $a$ and $\mu _{0}$
($a\geq 0$), the solitons are stable if the slope of the curve $N=N(\mu )$ ($N$ is
the total number of atoms) is negative, $\mathrm{d}N/\mathrm{d}\mu <0$. This is a
stability condition of the Vakhitov-Kolokolov (VK) type \cite{VK} (proof for the
case under consideration is given in section~\ref{stab}). At $\mu =\mu
_{\mathrm{max}}$, where the curves in Fig.~2 start from the zero number of atoms,
the solitons bifurcate from the trivial solution $U=V=0$. As $\mu $ approaches $\mu
_{\max }$, the width of the solitons tends to infinity, while their amplitudes
decrease to zero.

\begin{figure}[f]
\psfig{file=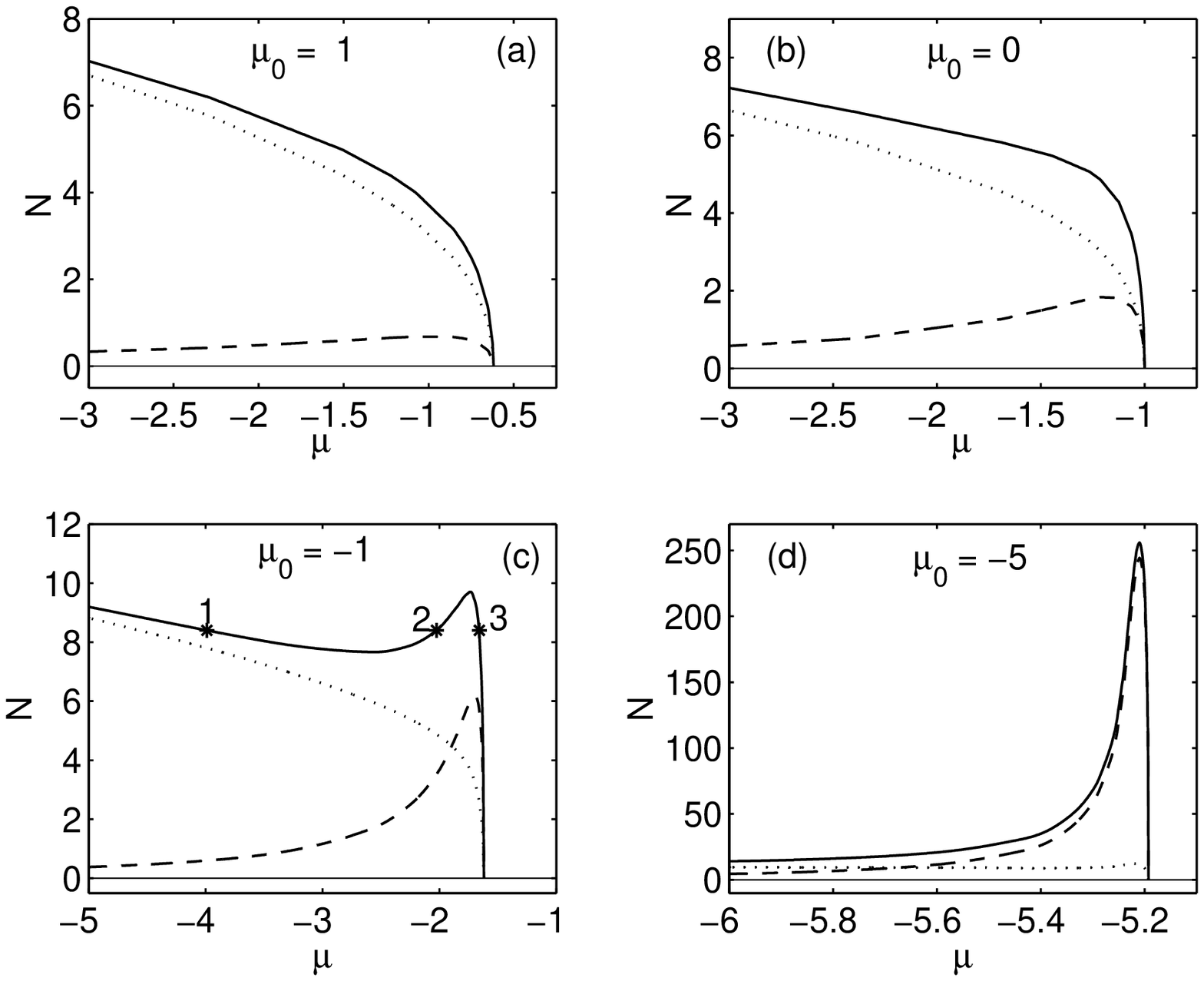,height=0.65\textwidth,width=0.65\textwidth} \caption{The
numbers of atoms in the two condensates vs. the chemical potential
$\protect\mu$ for $a=0$ and $\mu_0=1,\,0,\,-1,\,-5$. The total number of atoms
is given by solid curves, the number of atoms in the $u$-condensate by dotted
and  the number of atoms in the $v$-condensate by dashed curves. Here the
$v$-condensate contains non-interacting atoms, while the atomic interactions in
the $u$-condensate are attractive.}
\end{figure}

Decrease in the chemical potential difference $\mu _{0}$ causes the curve
$N=N(\mu )$ to sag in and develop a local maximum close to the upper limit
value $\mu=\mu _{\max }$, see Figs.~2(b)-(d) [in Fig.~2(d) we show only the
sagging part of the curve, the shape of the whole curve being similar  to that
in Fig. 2(c)]. It is interesting to note that for $\mu $ close to the value
that corresponds to the maximum of $N=N(\mu )$ almost all the atoms are trapped
in the linear $v$-condensate. The share of the number of atoms trapped in the
condensate with no atomic interaction grows even further (for $\mu$ around the
local maximum of $N=N(\mu)$) with further decrease of $\mu _{0}$ towards minus
infinity. For instance, for $a=0$ and $\mu _{0}=-10$, the number of atoms in
the linear condensate can be up to $99\%$ (the respective function $N=N(\mu )$
is similar to that in Fig.~2(d), but with the local maximum at $N\approx
2500$). On the contrary, for any given $\mu _{0}$,  for large negative values
of the chemical potential $\mu$ almost all the atoms are trapped in the
$u$-condensate.

The deformation of the solitons with the variation of the chemical potential
for fixed $a$ and $\mu _{0}$ is illustrated by Fig.~3. The soliton solutions
are plotted for three values of the chemical potential, which are marked by
stars in Fig.~2(c), that correspond to the same total number of atoms. Note
that, while at point~1, where $\mu =-3.9894$, the $u$-component of the soliton
is significantly higher, the $v$-component takes over as one moves to the right
along the curve $N=N(\mu )$ in Fig.~2(c): for $\mu =-2.0236$ and $\mu =-1.6576$
(the points 2 and 3, respectively) the $v$-component of the soliton solution
has a larger amplitude.

\begin{figure}[f]
\psfig{file=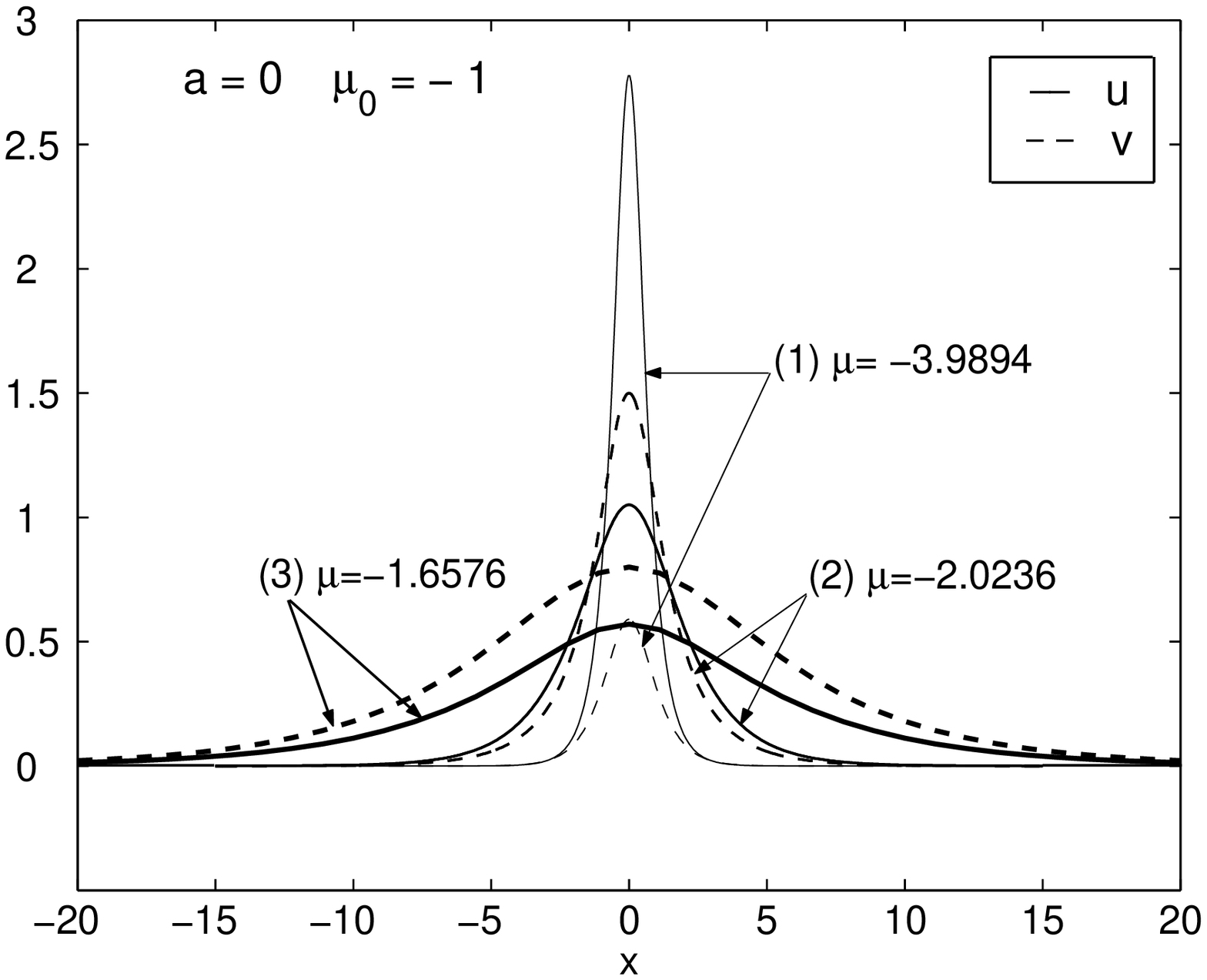,height=0.55\textwidth,width=0.55\textwidth} \caption{The
soliton solutions with a fixed total number of trapped atoms which correspond
to the three values of chemical potential marked by stars in Fig.~2(c). Solid
and dashed curves correspond to the $u$- and $v$- components of the solitons,
respectively. }
\end{figure}

\medskip \noindent \textsl{II. Small positive $a$}
\nopagebreak

\noindent For small positive values of $a$ (roughly, for $a\leq 0.1$) and positive
or small negative $\mu _{0}$, the numbers of atoms in the two condensates vs. the
chemical potential have the forms similar to those of Figs.~2(a) through (c). For
instance, for $a=0.1$ and for $\mu _{0}=-1$, the curves are found to be similar to
those in Fig.~2(c). Thus, for such values of $a$ and $\mu _{0}$, the solitons
bifurcate from the trivial solution at $\mu =\mu _{\max }$ (\ref{rangemu}), as in
the case of zero $a$. Moreover, for very small values of $a$ and some negative $\mu
_{0}$, almost all the atoms can be trapped in the condensate with the (weak)
repulsive inter-atomic interaction. As an example, a zoomed-in part of the
functions $N=N(\mu )$, $N_u=N_u(\mu)$ and $N_v=N_v(\mu)$ around the local maximum,
together with the soliton solutions, are displayed in Fig.~4 for $a=0.001$ and $\mu
_{0}=-5$. At point~2 in this figure, about $96\%$ of all the atoms are trapped in
the $v$-condensate with the weak repulsive interaction [a zoomed-in part of
Fig.~2(d) would be similar to left panel of Fig.~4 with a slightly lower local
maximum].

\begin{figure}[f]
\psfig{file=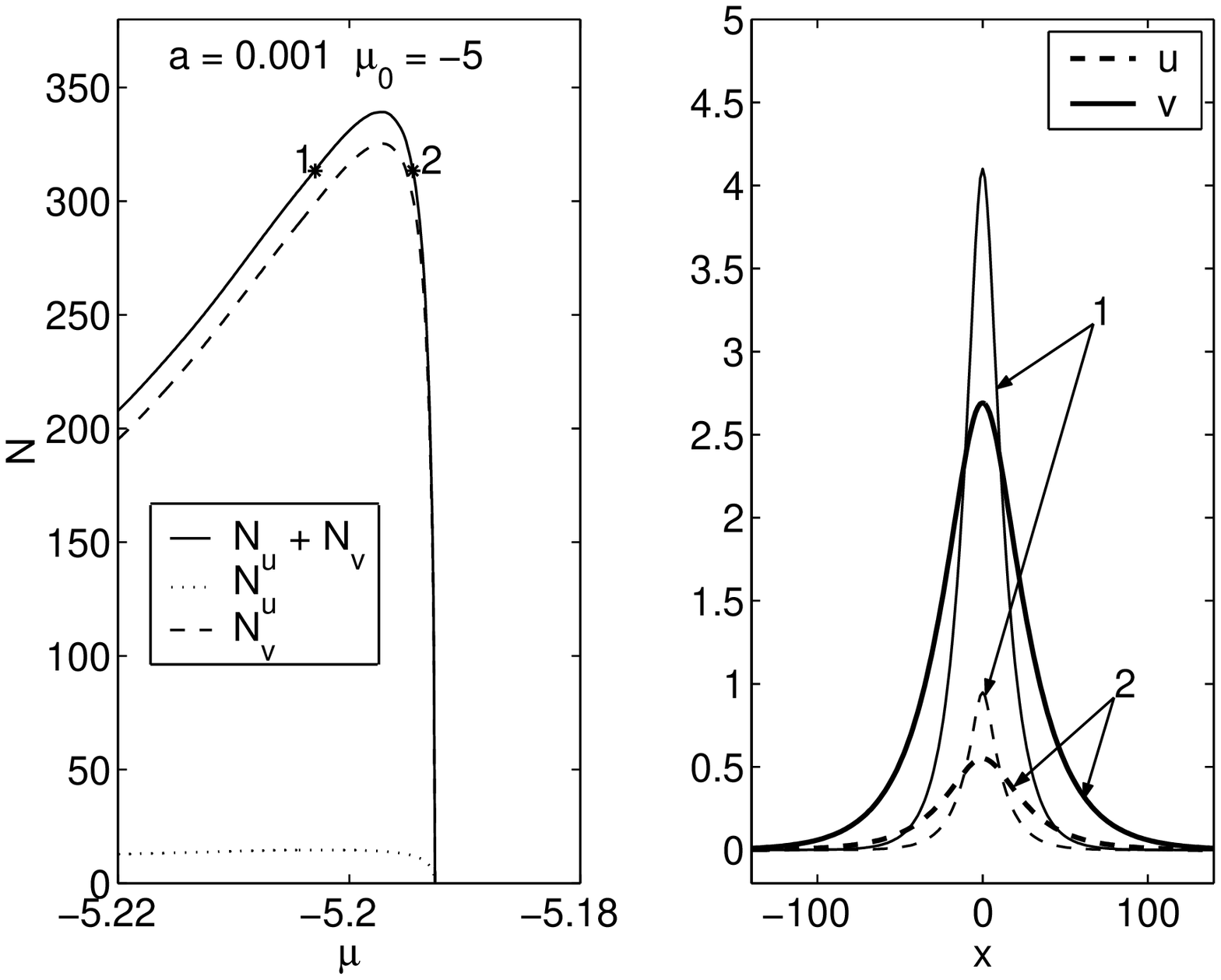,height=0.5\textwidth,width=0.65\textwidth} \caption{The
solitons with nearly all the atoms collected in the condensate with weak
repulsion ($a=0.001$). The left panel shows a zoomed-in part of the curves
$N=N(\mu )$, $N_u=N_u(\mu)$ and $N_v=N_v(\mu)$ about the local maximum. The
soliton solutions corresponding to the stars on the curve $N=N(\protect\mu)$
are given in the right panel (here the solid curves correspond to the
$v$-component, while dashed to the $u$-component).}
\end{figure}

However, the positive scattering length (self-repulsion) in the $v$-condensate
brings new features too, as compared to the linear $v$-condensate. Most
importantly, the grows of the local maximum of $N=N(\mu )$ with decrease of
$\mu_0$ eventually saturates and changes for decrease, i.e., the maximum
reaches its peak value at some negative $\mu _{0}$. Further decrease of $\mu_0$
causes the local maximum to disappear. Instead, a divergence develops  at the
boundary: $\mathrm{d}N/\mathrm{d}\mu\to\infty $ as $\mu\to\mu_{\mathrm{max}}$.
For instance, for $a=0.1$ and $\mu _{0}=-2$ there is no local maxima at all,
while the share of atoms in the $v$-condensate is large for $\mu $ close to
$\mu _{\max }$ due to the above mentioned divergence. Recall that the VK
stability criterion demands the negative slope of $\mathrm{d}N/\mathrm{d}\mu $,
hence the solitons with a large share of atoms in the repulsive condensate are
unstable for such values of $a$ and $\mu_0$. This case is illustrated by Fig.
5, where we give the numbers of atoms vs. chemical potential and typical $u$-
and $v$-shapes of the corresponding unstable soliton with larger number of
atoms in the repulsive condensate.

\begin{figure}[f]
\psfig{file=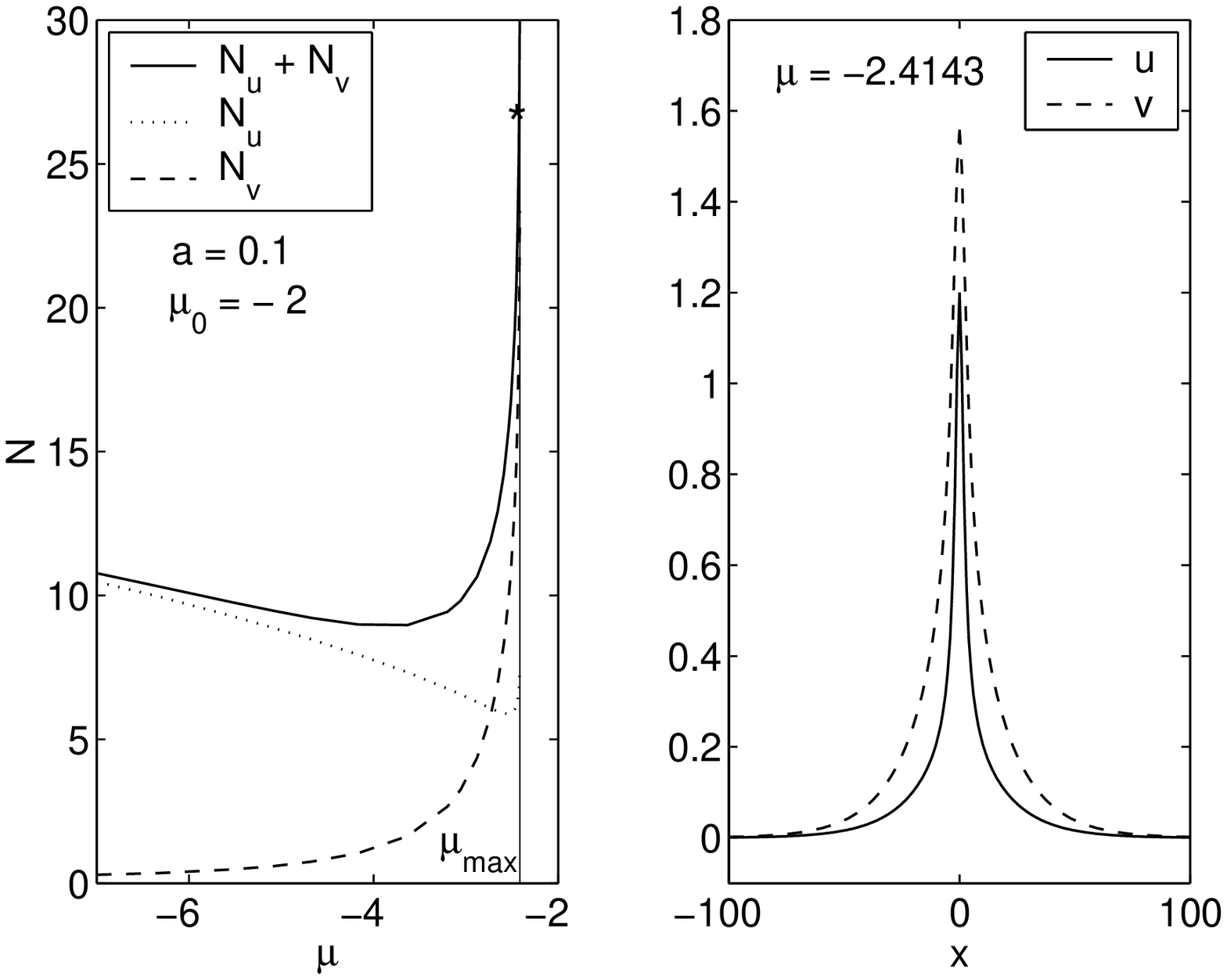,height=0.5\textwidth,width=0.65\textwidth} \caption{Typical
curves for the number of atoms vs. chemical potential (left), and an example of
the unstable soliton solution (right) featuring a large share of atoms in the
repulsive condensate.  Here $a=0.1$ and $\protect\mu_0=-2$. The soliton
corresponds to the star on the curve $N=N(\protect\mu)$ (the solid curve
corresponds to the $u$-component and dashed to the $v$-component).}
\end{figure}

The following property of the soliton solutions is observed as $\mu \rightarrow
\mu _{\max }$ (for curves similar to those in the left panel of figure~5): the
soliton's amplitude remains bounded, while its width is not. Therefore, our
conjecture is that the solitons develop a ``pedestal'' (long shelf) of an
increasing width, as the chemical potential approaches its limit value $\mu
_{\max }$ (this limit is very difficult to study numerically, precisely for the
same reason). In any case, all such solitons are unstable, since the slope of
the corresponding curve $N=N(\mu )$ is positive, see Fig.~5.

We have checked that, for various small positive $a$ and sufficiently large
negative values of $\mu _{0}$, the dependence of number of atoms on the
chemical potential is quite similar to what is shown in Fig.~5. Vice versa, for
arbitrary negative $\mu _{0}$ there is a threshold value of $a$ such that, for
$a$ larger  than this value, the derivative $\mathrm{d}N/\mathrm{d}\mu $ grows
to infinity as $\mu \rightarrow \mu _{\max }$. For example, for $\mu _{0}=-1$
it was found that, for $a=0.25$, the dependence of the numbers of atoms on the
chemical potential has essentially the same form as in Fig.~5.

The results on the solitons for small (positive) values of $a$ and various
$\mu_{0}$ can be summarized as follows: the soliton solutions with $\mu$ close
to $\mu_{\mathrm{max}}$ and positive or small negative $\mu _{0}$  are similar
to those displayed in Fig.~3, and for large negative $\mu _{0}$ are similar to
the soliton displayed in the right panel of Fig. 5. On the other hand, for
$\mu\ll \mu_{\mathrm{max}}$ the solitons are similar to the solution with
$\mu=-3.9894$ displayed in Fig.~3 (tagged by 1).

\medskip \noindent\textsl{III. Large positive $a$}

For $a\sim 1$ and negative values of $\mu _{0}$, the curves $N=N(\mu )$,
$N_{u}=N_{u}(\mu )$, and $N_{v}=N_{v}(\mu )$ are still similar to those in
Fig.~5, while for positive $\mu _{0}$ they are similar to the curves in
Fig.~2(a). Increasing $a$ further (i.e., making the repulsion between atoms in
the $v$-condensate still stronger), we have found that the numbers of atoms in
the condensates vs. the chemical potential take the form of what is displayed
in the left panel of Fig.~5 also for positive $\mu _{0}$.

The value separating the two different types of behavior of $N=N(\mu)$ as
$\mu\to\mu_{\mathrm{max}}$, i.e., given by Fig.~2(a) and the left panel of
Fig.~5, was found to be $a=1$. At $a=1$ and $\mu _{0}=0$, the soliton has its
amplitude gradually vanishing, as $\mu \rightarrow \mu _{\mathrm{\max}}$, but,
nevertheless, the number of atoms approaches a \emph{non-zero} value, which is
evident from the upper panel of Fig.~6. This seemingly strange bifurcation
implies that in the limit $\mu\to\mu_\mathrm{max}$ the soliton's width is
coupled to its amplitude, thus the integral that gives the number of atoms
remains constant (see the bottom panel in Fig.~6). Such a bifurcation may be
naturally called a discontinuous one, as the soliton bifurcates from the zero
solution without a visible jump in the shape, but with a jump in the number of
atoms.

\begin{figure}[f]
\psfig{file=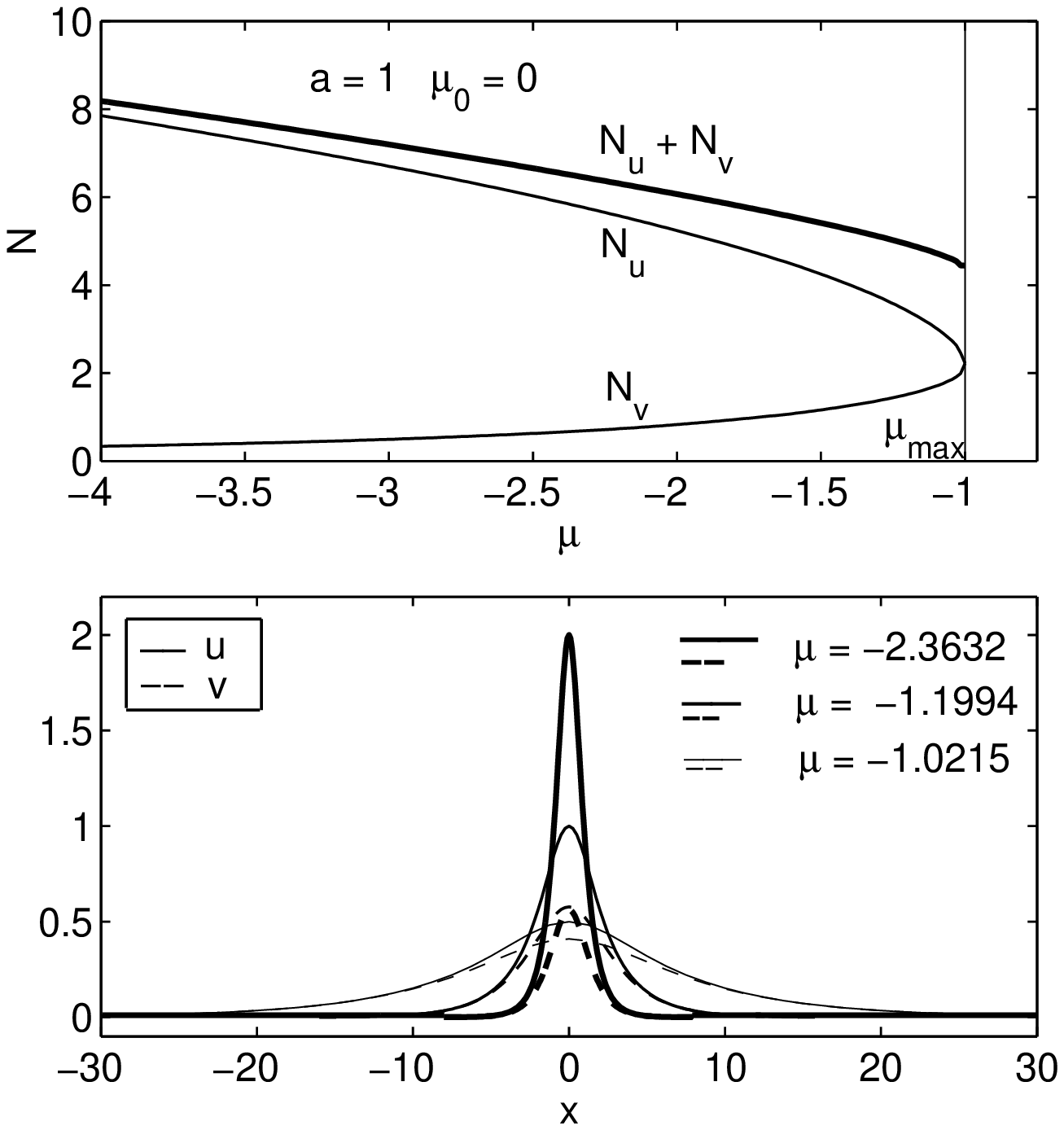,height=0.6\textwidth,width=0.6\textwidth} \caption{The
number of atoms in the condensates vs. the chemical potential (top), and
examples of the solitons (bottom) in the case when the soliton bifurcation from
the zero solution is accompanied by a jump in the number of atoms (here $a=1$
and $\mu_0=0$).}
\end{figure}

It may be interesting to check if the numerically found solitons for $a\ge 0$
are approximated by a sech-based ansatz. Let us assume, for example, that the
soliton solutions can be approximated by the
$\mathrm{sech}^\alpha\,$-functions, i.e., the solitons have the shape of $f =
A\mathrm{sech}^\alpha(x/d)$, where $A$ (amplitude) and $d$ (width), in general,
different for the two components,  are the dynamic parameters (determined
through the variational analysis) and the power $\alpha$ is fixed. This is an
improvement of the usual sech-type ansatz. Given a numerical soliton solution
we need to determine $\alpha$ such that the corresponding
sech$^\alpha$-function approximates the soliton in some sense. This can be done
in many different ways. Here we adopt as the proximity measure the following
functional
\begin{equation}
P_f = \frac{\int\limits_{-\infty }^{\infty }\mathrm{d}x\,x^{2}f\int\limits_{-
\infty }^{\infty }\mathrm{d}x\,f^{2}}{\int\limits_{-\infty }^{\infty
}\mathrm{d} x\,f\int\limits_{-\infty }^{\infty }\mathrm{d}x\,x^{2}f^{2}}\,.
\label{condit_num}
\end{equation}
Inverting the function $P=P_{\mathrm{sech}^\alpha}\equiv P(\alpha)$ and using
the numerically computed $P_S$ (for a numerically found soliton solution
$S=S(x)$) we get the corresponding $\alpha$: $\alpha = \alpha(P_S)$. [The
proximity measure was chosen in the ad hoc way. However, there are two
conditions to be satisfied: ($i$) the functional must be independent of both
$A$ and $d$; ($ii$) the function $P=P(\alpha)$ must be monotonous (it is
decreasing for $P_f$).] If the solitons could be approximated by the above
ansatz, the expected result would be a sharply peaked distribution of the power
parameter values $\alpha$ computed for the numerically found solitons. Then one
could assume the value of $\alpha$ corresponding to the peak as the
approximation value.

We have computed the power parameter $\alpha$ in the above approximation for
the numerically found soliton solutions for various $a$, $\mu _{0}$ and $\mu $.
For instance, for the solitons shown in Fig.~3 the corresponding values of
$\alpha$ are as follows (the subscripts refer to the $u$- and $v$-components of
the soliton)
\[
\begin{array}{cc}
\alpha_{u}^{(1)}=0.825, & \alpha_{v}^{(1)}=1.61, \\
\alpha_{u}^{(2)}=0.218, & \alpha_{v}^{(2)}=1.01, \\
\alpha_{u}^{(3)}=0.711, & \alpha_{v}^{(3)}=0.922.
\end{array}
\]
Evidently the soliton solutions  cannot be approximated by the
sech$^\alpha$-type functions with the same common power $\alpha$ due to a broad
distribution of the power parameters corresponding to the solitons.  Here we
note that, in contrast, the variational approach based on the sech-type ansatz
with the amplitudes and widths as the dynamic variables was successfully
applied before to the solitons for $a<0$ (see Refs.~\cite{M1,K1,K2} where the
case of $a=-1$ was considered).

\subsection{ Solitons for $a<0$}
\label{a<0}

In the case of $a<0$, when the interactions are attractive in both condensates,
it is enough to consider the interval $-1\leq a<0$, as Eqs.
(\ref{E4U})-(\ref{E4V}) are invariant against the following substitution:
\begin{equation}
U\rightarrow \tilde{U}=\sqrt{-a}V,\quad V\rightarrow \tilde{V}=\sqrt{-a}
U,\quad \mu _{0}\rightarrow \tilde{\mu}_{0}=-\mu _{0},\quad a\rightarrow
\tilde{a}=\frac{1}{a},  \label{invar}
\end{equation}
and $\mu \rightarrow $ $\tilde{\mu}=\mu -\mu _{0}$.

The solitons for $a<0$ were  studied before in the context of the nonlinear
dual-core optical fibers \cite{A1,A2,A3,M1,K1,K2}, chiefly in the particular
case $a=-1$. Though our new results, presented below, pertain to $a\neq -1$, we
also consider in detail the previously studied case of $a=-1$  and obtain some
results, which were not known before.

We will frequently refer to the special case when $a$ and $\mu _{0}$ are
related as follows $\mu _{0}=\sqrt{-a}-{1}/\sqrt{-a} \equiv \mu _{0}\left(
a\right)$. Note that only in this case Eqs.~(\ref{E4U})-(\ref{E4V}) admit
(real) sech-type soliton solutions.

In section~\ref{sec1} it was shown that there are two branches of the in-phase
solitons, one given by Eq. (\ref{exactinph}) for arbitrary (negative) $a$ and
$\mu _{0}=\mu _{0}(a)$, and the other one bifurcating from these sech-type
solitons at $\mu =\mu _{\mathrm{bif}}$, see Eq.~(\ref{bif}). We have found
numerically that these branches \emph{ coexist} for arbitrary values of $a$ and
$\mu _{0}$, although they do not always intersect, hence do not always undergo
a collision bifurcation. Thus, the sech-type soliton (\ref{exactinph}) is a
special case of a broader family of soliton solutions. Moreover, we have
verified numerically that this broad family of (in-phase) solitons corresponds
to the $k_{1}$-branch of the dispersion relation, see Eq. (\ref{disp}), i.e.
the solitons exist for $\mu $ satisfying the condition (\ref{rangemu}).

To present the results in more detail, we first consider small (negative)
values of $a$. We note that the bifurcation involving the two branches of
solitons, with one branch corresponding to the sech-type solitons
(\ref{exactinph}), belongs to the \textit{tangential} type, i.e., it is a
one-sided cusp bifurcation (consult, for instance, Ref.~\cite{Iooss}). This is
clearly seen from Fig.~7, where we take the values $a=-0.1$ and $\mu _{0}=\mu
_{0}(a)$ (this statement is also proven analytically in section~\ref{stab}).
For small deviations $\mu _{0}-\mu _{0}(a)$, and small $a$ (roughly for
$-0.5\leq a<0$), the bifurcation is similar. For larger deviations $\mu
_{0}-\mu _{0}(a)$, the two branches develop a trend to collide, which gives
rise to a picture resembling a collision bifurcation (see Fig.~8, where this
type of behavior is shown, though for a larger negative value of $a$).

\begin{figure}[f]
\psfig{file=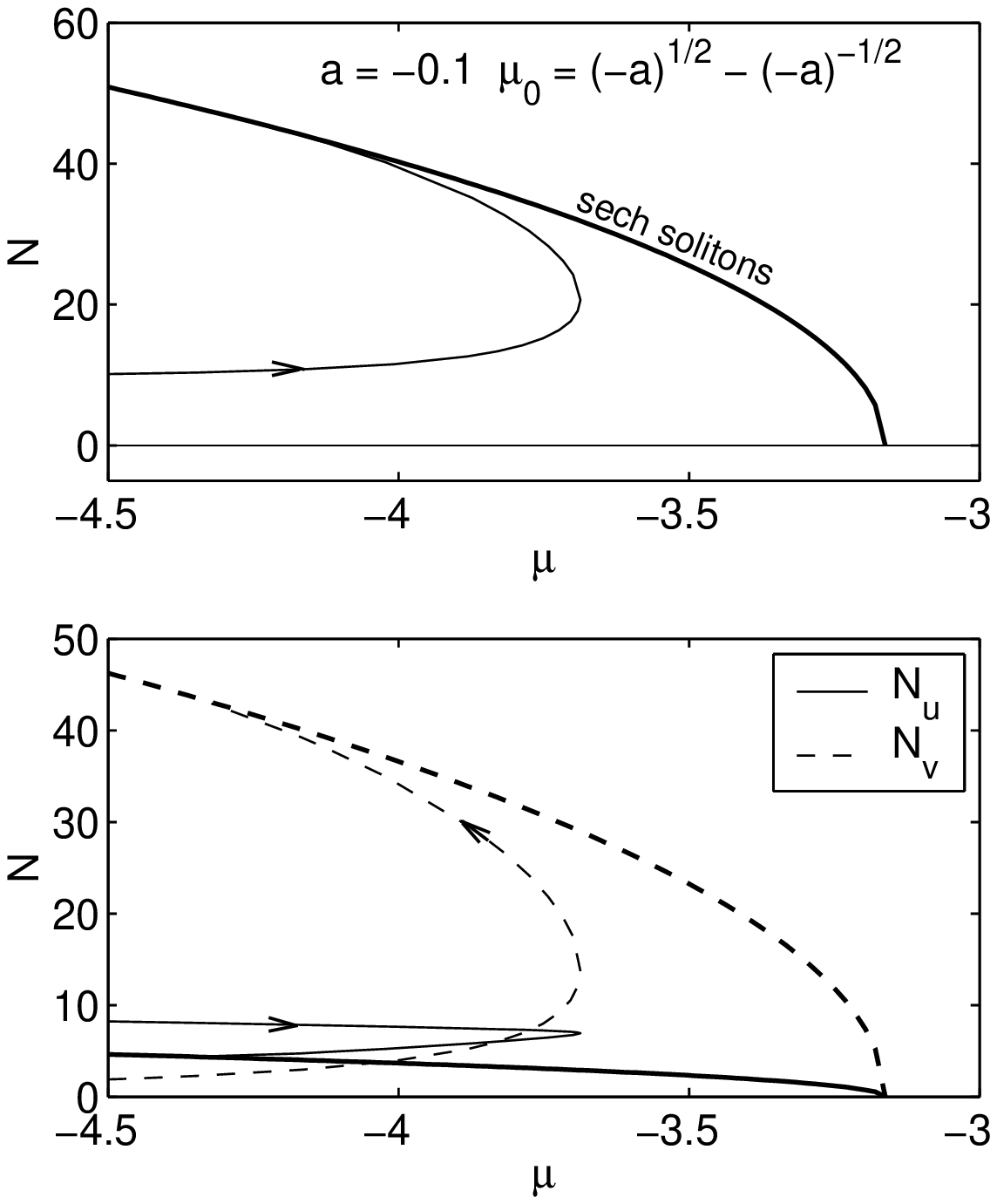,height=0.6\textwidth,width=0.6\textwidth} \caption{The
bifurcation diagram for $a=-0.1$ and $\protect\mu_0 = \protect
\sqrt{-a}-1/\protect\sqrt{-a}\approx -2.846$. The top panel shows the total
numbers of atoms in the condensates for the two branches of the soliton
solutions. The bottom plot shows the numbers of atoms in each of the two
condensates. In the figure, the thick curves correspond to the family of the
sech-type solitons. }
\end{figure}

Finally, for small negative $a$ and for small but finite deviations $\mu
_{0}-\mu _{0}(a)$ we have checked that the solitons belonging to the branch
which corresponds to the sech-type solitons at $\mu _{0}=\mu _{0}(a)$ can be
uniformly approximated by the sech ansatz for \emph{all} values of the chemical
potential $\mu$.  This is, however, not so for larger negative values of $a$
(see the discussion below for $a=-1$).

We have found that the collision bifurcation is replaced by the coexisting
branches when the deviation  $\mu_{0}-\mu_{0}(a)$ has large enough modulus. For
instance, when $a=-0.8$, the two branches are clearly separated for $\mu
_{0}=-0.2$, see Fig.~8, which is $\simeq 10\%$ away from the corresponding
value $\mu _{0}(a=-0.8)\approx -0.22$ (the case of the small deviation $\mu
_{0}-\mu _{0}(a)$ is analyzed in detail for $a=-1$ below).

\begin{figure}[f]
\psfig{file=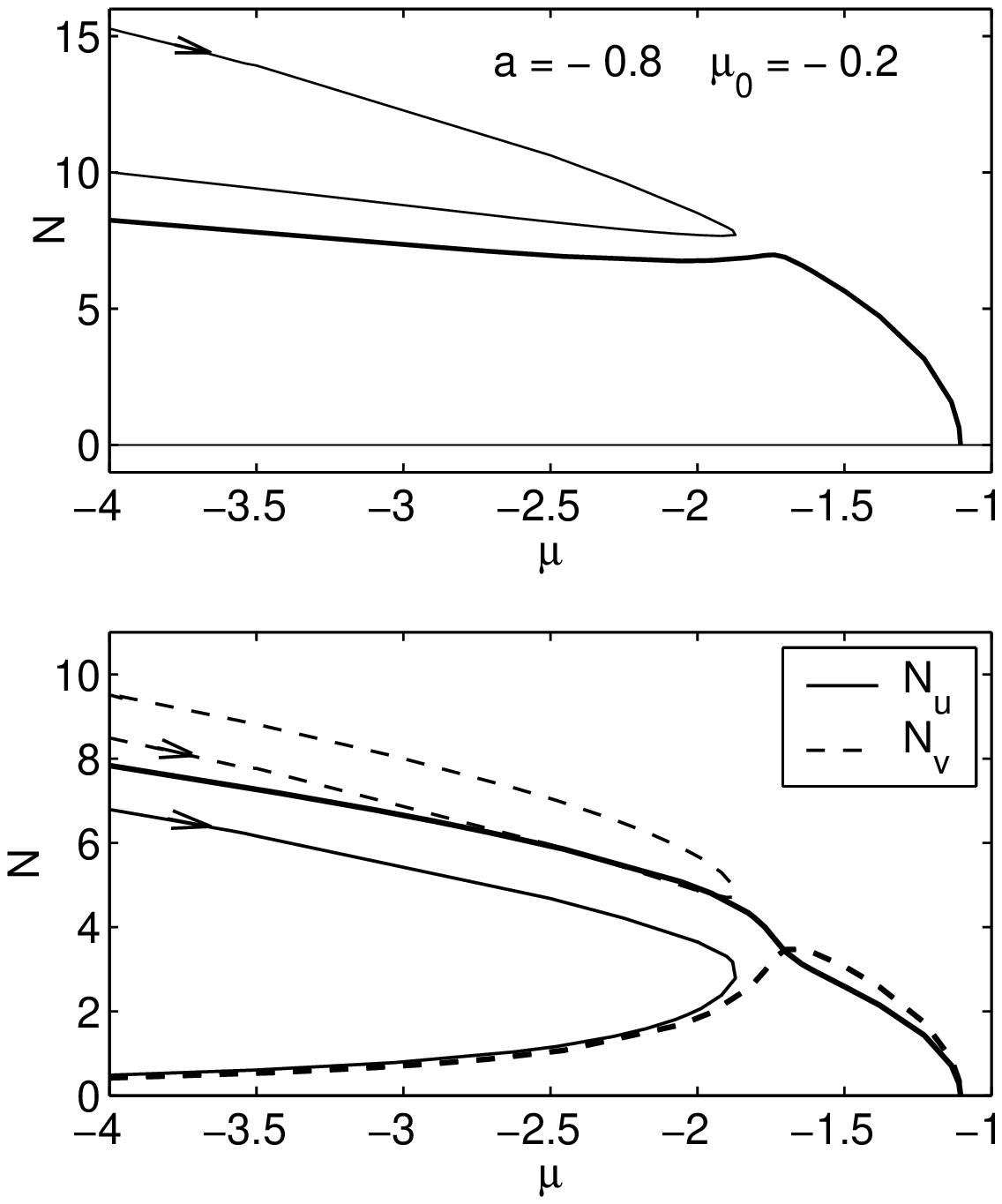,height=0.6\textwidth,width=0.6\textwidth} \caption{The same
as in Fig. 7 for $a=-0.8$ and $\protect\mu_0 = -0.2$, except that here
$\mu_0\ne \mu_0(a)$.}
\end{figure}

In Fig.~9 we display the bifurcation diagrams for $a=-1$. Here it is worth
noting that, due to the symmetry against the substitution (\ref {invar}) the
solution components $U=U(x)$ and $V=V(x)$ are interchangeable in this case. For
instance, the sech-type solitons have identical components, $U=V$ (the
symmetric solitons), while the solitons belonging to the other branch (the
asymmetric ones) admit two possibilities: $U>V$, or $V>U$. Only the diagrams
with $U>V$ are displayed in Figs.~9(b) and (d).

\begin{figure}[f]
\psfig{file=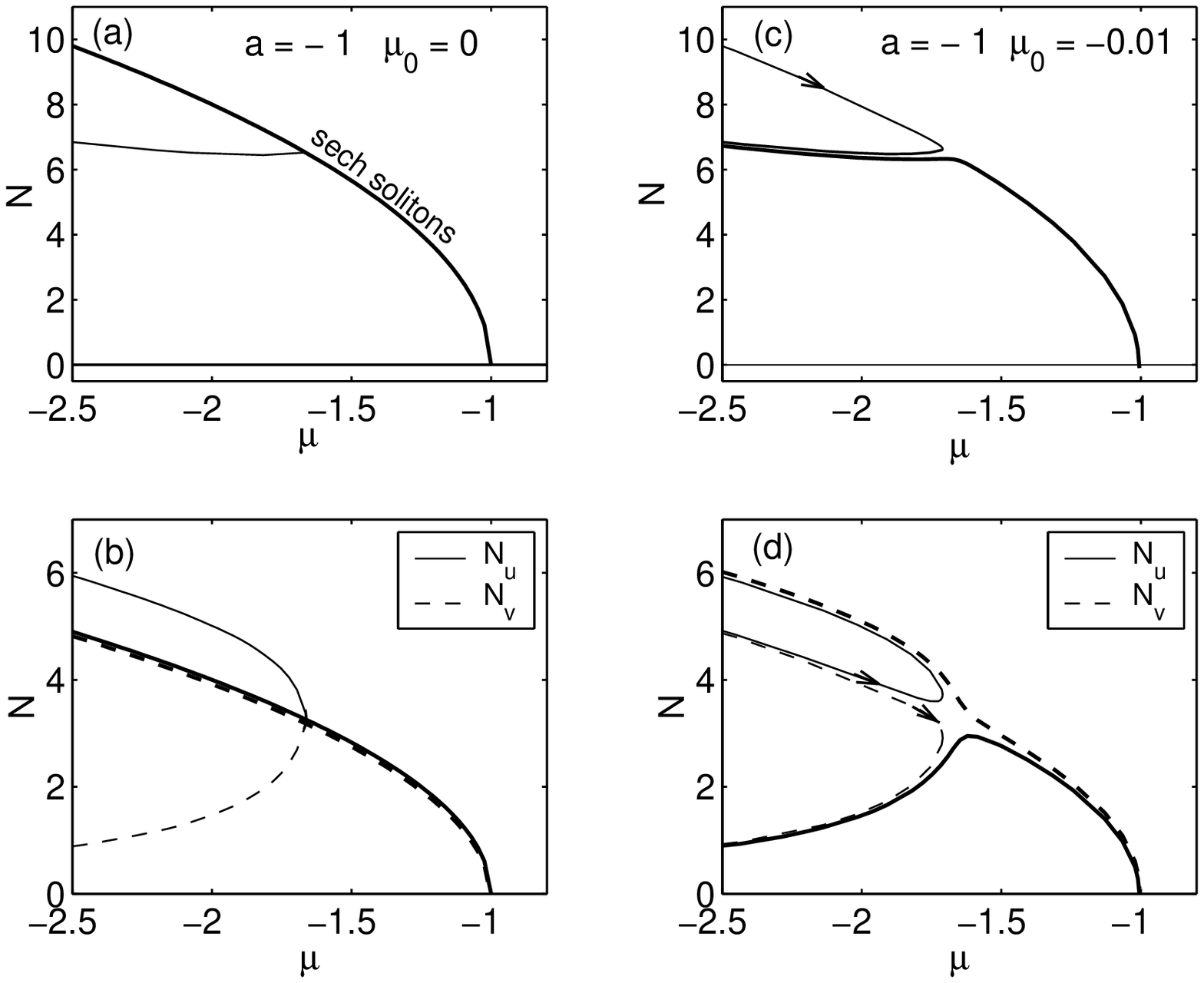,height=0.6\textwidth,width=0.6\textwidth} \caption{The
panels (a) and (b) show the bifurcation diagrams for $a=-1$ and $\protect\mu_0
= 0$, and the panels (c) and (d) show the diagrams for $\protect\mu_0 = -0.01$.
As well as in Figs. 7 and 8, the top and bottom plots display, respectively,
the total number of atoms, and the numbers of atoms in each condensate. The
thick curves correspond to the sech-type branch of solitons and its
deformations.}
\end{figure}

First of all, we notice that Figs. 9(a) and 9(b) reproduce the known
bifurcation diagram for $\mu_0=0$ \cite{A1} (note that in this case
$\mu_{0}=\mu _{0}(a=-1)$). Although it is demonstrated in section~\ref{stab}
that the asymmetric solitons bifurcate tangentially from the symmetric branch,
we were not able to show this feature in the picture, because a region where it
takes place for $a=-1$ and $\mu _{0}=0$ is extremely narrow (for the same
reason, this fact was left unnoticed in Ref.~\cite{A1}).

As soon as the chemical-potential difference $\mu _{0}$ deviates from zero (in
terms of the dual-core-fiber model, this corresponds to a phase mismatch
between the cores), the collision bifurcation undergoes a significant
transformation, see figure~9(c). It is interesting to note that, as soon as
$\mu _{0}$ goes away from zero (we take $\mu _{0}=-0.01$), the former sech-type
branch divides itself into two different branches of  soliton solutions, while
the former asymmetric branch splits into two close curves, which pertain to
different branches as well.

Consider the top part of the branch with the arrow in Fig.~9(c). In this
region, the power parameters $\alpha_{u}$ and $\alpha_{v}$, numerically
computed using Eq.~(\ref{condit_num}), differ from $1$ by less than
$2\%$\thinspace. Although this curve corresponds to the solitons which are well
approximated by the sech ansatz and the corresponding soliton solutions have
nearly identical components, $U\approx V$, it nevertheless can be identified
with the deformed asymmetric branch of solitons. This statement is supported by
comparison of Fig. 9(d) with Fig. 9(b): in both cases, the bifurcation diagram
has a pitchfork-type form, but in the case of Fig. 9(d) the stem and the middle
prong of the pitchfork, both lying very close to the sech-type branch for $\mu
_{0}=0$, belong to different branches of  solutions. We were not able to trace
the bifurcation point $\mu _{\mathrm{bif}}$ by continuing Fig.~9(c) in the
direction of negative $\mu$, i.e., for  $\mu \rightarrow -\infty$, and checking
stability of the solitons (the bifurcation point is the threshold of the
soliton instability, see section~\ref{stab}). Thus, when $a=-1$, the collision
bifurcation is replaced by two coexisting branches of solitons already for $\mu
_{0}=-0.01$.

Finally, we observe that the branch of the soliton solutions denoted by the
thin curves in Figs.~7 through 9, which corresponds to the asymmetric solitons
for $a=-1$ and $\mu _{0}=0$, undergoes a turning-point bifurcation. The turning
point is clearly seen in these figures.

As it was mentioned above, we would not consider bifurcations of the
out-of-phase (e.g., antisymmetric) solutions, as they are always unstable (see
also Ref. \cite{A2}). For the same reason, we do not consider of higher-order
multi-humped solitons, which are also found in the numerical solution, but turn
out to be unstable in all the cases.

\section{Soliton stability and evolution under perturbations}

\label{stab}

After having found the soliton solutions, the next necessary step is to analyze
the soliton stability under action of perturbations, as well as the fate of
unstable solitons. Accordingly, we split this section into two parts. In the
first part, we prove the above-mentioned VK stability criterion, $\mathrm{d}N/
\mathrm{d}\mu <0$, for the in-phase  soliton solutions of Eqs. (\ref
{E4U})-(\ref{E4V}) with $a\ge0$ and show where the VK criterion applies to the
solitons for $a<0$. For instance,  we demonstrate that the bifurcation point
(\ref{bif}), discussed in the previous section, is (quite naturally) an
instability threshold for the sech-type solitons (\ref {exactinph}). Our proof
of the stability criterion for the soliton solutions of the system
(\ref{E1})-(\ref{E2}) generalizes a similar proof for the solitons of a single
(scalar) equation, which can be found in Refs.~\cite{VK,Zakh}. In the second
part of this section, we numerically study the evolution of unstable solitons
subject to initial perturbations.

Here we point that for $a=-1$ (with $\mu_0=0$) the sech-type soliton solutions
(\ref{exactinph}) discussed in the present paper, and the in-phase solitons in
general, reduce to the solitons which were studied in Ref.~\cite{A1}, in the
context of the dual-core optical fibers. The stability properties of the
in-phase solitons for $a=-1$ were investigated in Ref.~\cite{A2}, and evolution
of unstable solitons under perturbations was simulated in Ref.~\cite{A3}.
Below, we (in particular) generalize the stability results of Ref.~\cite{A2}
for arbitrary $a<0$ and $\mu _{0}$, which is relevant to the model
(\ref{E1})-(\ref{E2}) for trapped BECs.

\subsection{Stability analysis}

\noindent \textsl{I. Stability of the in-phase solitons for $a\ge0$}

The soliton stability with respect to small perturbations is determined by the
system (\ref{E1})-(\ref{E2}) linearized about the soliton solution. Thus we
need to consider a perturbed soliton solution:
\[
u=e^{-i\mu t}\left[ U(x)+u_{1}(t,x)\right] ,\quad v=e^{-i\mu t}\left[
V(x)+v_{1}(t,x)\right] ,
\]
where $U(x)$ and $V(x)$ are the stationary soliton solution components, while
$u_{1}(t,x)\equiv \mathcal{U}_{R}(t,x)+i\mathcal{U}_{I}(t,x)$ and
$v_{1}(t,x)\equiv \mathcal{V}_{R}(t,x)+i\mathcal{V}_{I}(t,x)$ represent the
perturbation. Linearizing equations (\ref{E1})-(\ref{E2}) with respect to the
perturbation we arrive at the system
\begin{equation}
\partial _{t}\left(
\begin{array}{c}
\mathcal{U}_{R} \\
\mathcal{U}_{I} \\
\mathcal{V}_{R} \\
\mathcal{V_{I}}
\end{array}
\right) =\left(
\begin{array}{cccc}
0 & L_{0}^{(u)} & 0 & -1 \\
-L_{1}^{(u)} & 0 & 1 & 0 \\
0 & -1 & 0 & L_{0}^{(v)} \\
1 & 0 & -L_{1}^{(v)} & 0
\end{array}
\right) \left(
\begin{array}{c}
\mathcal{U}_{R} \\
\mathcal{U}_{I} \\
\mathcal{V}_{R} \\
\mathcal{V_{I}}
\end{array}
\right) \,,  \label{linprob}
\end{equation}
where the linear operators are defined as follows:
\[
L_{0}^{(u)}\equiv -\left( \frac{\mathrm{d}^{2}}{\mathrm{d}x^{2}}+\mu
+U^{2}(x)\right),\; L_{0}^{(v)}=-\left(
\frac{\mathrm{d}^{2}}{\mathrm{d}x^{2}}+\mu -\mu _{0}-aV^{2}(x)\right),
\]
\begin{equation}
L_{1}^{(u)}\equiv -\left( \frac{\mathrm{d}^{2}}{\mathrm{d }x^{2}}+\mu
+3U^{2}(x)\right),\; L_{1}^{(v)}=-\left( \frac{\mathrm{d}^{2}}{\mathrm{d}
x^{2}}+\mu -\mu _{0}-3aV^{2}(x)\right).
\label{opers}
\end{equation}
The solitons are unstable if the matrix operator on the right-hand side of
equation (\ref{linprob}) has an eigenvalue $\lambda $ with the positive real
part [in fact, the eigenvalues are real, since $\Lambda _{0}$ is non-negative,
see Eqs. (\ref{eigenprob})-(\ref{operlam}) and the discussion below]. The
system (\ref{linprob}) can be written in an equivalent second-order form, hence
the corresponding eigenvalue problem can be reformulated for a fourth-order
differential operator acting on a two-component vector $(X_{u},X_{v})$, with
$X_{u}$ and $X_{v}$ defined as follows (consult also appendix~\ref{the mode})
\[
\mathcal{U}_{R}(t,x)=e^{-i\Omega t}X_{u}(x)+\mathrm{c.c.},\quad \mathcal{V}
_{R}(t,x)=e^{-i\Omega t}X_{v}(x)+\mathrm{c.c.}.
\]
Here, $\Omega =i\lambda $ has the meaning of a frequency. The linear eigenvalue
problem then takes the form
\begin{equation}
\Lambda _{0}\Lambda _{1}\left(
\begin{array}{c}
X_{u} \\
X_{v}
\end{array}
\right) =\Omega ^{2}\left(
\begin{array}{c}
X_{u} \\
X_{v}
\end{array}
\right) ,  \label{eigenprob}
\end{equation}
where we have introduced two symmetric matrix operators:
\begin{equation}
\Lambda _{0}\equiv \left(
\begin{array}{cc}
L_{0}^{(u)} & -1 \\
-1 & L_{0}^{(v)}
\end{array}
\right) ,\quad \Lambda _{1}\equiv \left(
\begin{array}{cc}
L_{1}^{(u)} & -1 \\
-1 & L_{1}^{(v)}
\end{array}
\right) .  \label{operlam}
\end{equation}

The proof of the stability criterion relies on properties of the factorization
operators $\Lambda _{0}$ and $\Lambda _{1}$. We first aim to show that $\Lambda
_{0}$ is non-negative, i.e., the scalar product $\langle \Psi |\Lambda _{0}|\Psi
\rangle $, defined as $\int_{-\infty }^{+\infty }\mathrm{d}x\Psi
^{\dagger}(x)\Lambda _{0}\Psi (x)$ (here $\dagger$ stands for the Hermitian
conjugation), is non-negative for any real two-component vector $\Psi (x)=(\psi
_{1}(x),\psi _{2}(x))^{T}$. Indeed, the following inequalities follow from the
definitions the operators $L_{0}^{(u)}$ and $L_{0}^{(v)}$:
\begin{equation}
 \int \mathrm{d}x\,\psi^* (x)L_{0}^{(u)}\psi (x)\geq \int
\mathrm{d}x\,\frac{V(x)}{ U(x)}|\psi(x)| ^{2},\; \int \mathrm{d}x\,\psi^*
(x)L_{0}^{(v)}\psi (x)\geq \int \mathrm{d}x\,\frac{U(x)}{V(x)}|\psi(x)|^{2}.\;
\label{ineqL0}
\end{equation}
To arrive at Eq.~(\ref{ineqL0}) it is enough to note that, due to Eqs.
(\ref{E4U})-(\ref{E4V}), these operators can be rewritten as
\[
L_{0}^{(u)}=-\frac{1}{U(x)}\frac{\mathrm{d}}{\mathrm{d}x}U^{2}(x)\frac{\mathrm{d}}{
\mathrm{d}x}\frac{1}{U(x)}+\frac{V(x)}{U(x)},\;  L_{0}^{(v)}=-\frac{1}{V(x)}
\frac{\mathrm{d}}{\mathrm{d}x}V^{2}(x)\frac{\mathrm{d}}{\mathrm{d}x}\frac{1}{V(x)}+
\frac{U(x)}{V(x)}.
\]
Taking advantage of the inequalities (\ref{ineqL0}) and of the fact that, for
the in-phase soliton solutions, the product $U(x)V(x)$ is positive, we get
\[
\langle \Psi |\Lambda _{0}|\Psi \rangle =\int \mathrm{d}x\,\left\{
\psi^*_{1}(x)L_{0}^{(u)}\psi _{1}(x)+\psi^*_{2}(x)L_{0}^{(v)}\psi _{2}(x)
-\psi^*_{1}(x)\psi _{2}(x) - \psi_{1}(x)\psi^*_{2}(x)\right\}
\]
\[
\geq \int \mathrm{d}x\,\left\{ \frac{V(x)}{U(x)}|\psi _{1}(x)|^2+\frac{U(x)}{
V(x)}|\psi _{2}(x)|^2 -\psi^*_{1}(x)\psi _{2}(x) - \psi_{1}(x)\psi^*_{2}(x)\right\}
\]
\[
=\int \mathrm{d}x\,\left| \sqrt{\frac{V(x)}{U(x)}}\psi _{1}(x)-\sqrt{\frac{U(x)
}{V(x)}}\psi _{2}(x)\right| ^{2}\geq 0.
\]

The operator $\Lambda _{0}$ has, in fact, a zero eigenvalue, with the soliton
solution proper $(U,V)^{T}$ being the corresponding eigenfunction:
\begin{equation}
\Lambda _{0}\left(
\begin{array}{c}
U(x) \\
V(x)
\end{array}
\right) =0.  \label{zeroL0}
\end{equation}

In its turn, the other factorization operator $\Lambda _{1}$ also has a zero
mode,
\begin{equation}
\Lambda _{1}\left(\begin{array}{c}
{\mathrm{d}U(x)}/{\mathrm{d}x} \\
{\mathrm{d}V(x)}/{\mathrm{d}x}
\end{array}
\right) =0,  \label{zeroL1}
\end{equation}
which, however, has a node (zero) at $x=0$. Thus,  according to the Sturm
oscillation theorem, the operator $\Lambda _{1}$ may have negative eigenvalues (it
is found numerically that $\Lambda_1$ has either one or two negative  eigenvalues).
In general, it may have two nodeless eigenfunctions corresponding to two negative
eigenvalues:
\[
\Lambda _{1}\left(
\begin{array}{c}
f^{(1)}(x) \\
g^{(1)}(x)
\end{array}
\right) =\lambda _{1}\left(
\begin{array}{c}
f^{(1)}(x) \\
g^{(1)}(x)
\end{array}
\right) ,\quad \Lambda _{1}\left(
\begin{array}{c}
f^{(2)}(x) \\
g^{(2)}(x)
\end{array}
\right) =\lambda _{2}\left(
\begin{array}{c}
f^{(2)}(x) \\
g^{(2)}(x)
\end{array}
\right) ,
\]
where $f^{(1)}g^{(1)}>0$ and $f^{(2)}g^{(2)}<0$. However, it is shown in
appendix \ref{App} that $\Lambda_1$ cannot have an eigenfunction of the second
type for $a\geq 0$ (corresponding to a negative eigenvalue).

Return now to the eigenvalue problem (\ref{eigenprob}). We have
\begin{equation}
\Omega ^{2}\Lambda _{0}^{-1}\left(
\begin{array}{c}
X_{u} \\
X_{v}
\end{array}
\right) =\Lambda _{1}\left(
\begin{array}{c}
X_{u} \\
X_{v}
\end{array}
\right) + C_0\left(
\begin{array}{c}
U \\
V
\end{array}
\right) ,  \label{inverse}
\end{equation}
where $C_0 $ is a scalar constant. In deriving (\ref{inverse}), we have taken
into account that for \mbox{$\Psi=(X_u,X_v)^T$}
\begin{equation}
\langle (U,V)|\Psi\rangle =0,  \label{constr}
\end{equation}
since $\Lambda _{0}$ is a symmetric operator. Thus, the relation
(\ref{inverse}) is justified. The minimum eigenfrequency $\Omega _{0}$ can be
also found from the following minimization problem:
\begin{equation}
\Omega _{0}^{2}=\mathrm{min}\frac{\langle \Psi |\Lambda _{1}|\Psi \rangle }{
\langle \Psi |\Lambda _{0}^{-1}|\Psi \rangle }\,,  \label{minOmega}
\end{equation}
with $\Psi =(\psi _{1},\psi _{2})^{T}$ subject to the constraint (\ref
{constr}). To evaluate the sign of $\Omega _{0}^{2}$, one has to evaluate the
sign of the numerator in the expression (\ref{minOmega}). Due to the constraint
(\ref{constr}), the latter problem is equivalent to evaluation of the sign of
the lowest eigenvalue  of the following generalized eigenvalue problem,
\begin{equation}
\Lambda _{1}\left(
\begin{array}{c}
X \\
Y
\end{array}
\right) =\lambda \left(
\begin{array}{c}
X \\
Y
\end{array}
\right) +\beta \left(
\begin{array}{c}
U \\
V
\end{array}
\right),  \label{geneig}
\end{equation}
where $\beta $ is a constant and $\Psi=(X,Y)^{T}$ obeys Eq. (\ref{constr}).
From the above consideration, we know that $\Lambda _{1}$ has only one negative
eigenvalue, and the eigenfunction cannot be orthogonal to $(U,V)$ for $UV>0$,
since $f^{(1)}g^{(1)}$ is positive too. Therefore, the expansion of $(X,Y)^{T}$
over the eigenfunctions of $\Lambda _{1}$,
\[
\Lambda _{1}\left(
\begin{array}{c}
\phi _{1}^{(n)} \\
\phi _{2}^{(n)}
\end{array}
\right) =\lambda _{n}\left(
\begin{array}{c}
\phi _{1}^{(n)} \\
\phi _{2}^{(n)}
\end{array}
\right),
\]
has the following form (for simplicity of the presentation, the summation below
also denotes the integration over the continuous spectrum)
\begin{equation}
\left(
\begin{array}{c}
X \\
Y
\end{array}
\right) =\beta \sum_{n,n\ne2}{}\frac{\langle (\phi _{1}^{(n)},\phi
_{2}^{(n)})|\left(
\begin{array}{c}
U \\
V
\end{array}
\right) \rangle }{\lambda _{n}-\lambda }\left(
\begin{array}{c}
\phi _{1}^{(n)} \\
\phi _{2}^{(n)}
\end{array}
\right) ,  \label{XY}
\end{equation}
where the zero eigenvalue $\lambda _{2}=0$ does not enter the sum. In deriving
Eq. (\ref{XY}), we have made use of Eq. (\ref {geneig}) and the orthogonality
of the eigenfunctions of $\Lambda_1$. The constraint (\ref{constr}) requires
that the generalized eigenvalue $\lambda $ from Eq. (\ref{geneig}) satisfies
\begin{equation}
\beta \sum_{n,n\ne2}\frac{\langle (\phi _{1}^{(n)},\phi _{2}^{(n)})|\left(
\begin{array}{c}
U \\
V
\end{array}
\right) \rangle ^{2}}{\lambda _{n}-\lambda }\equiv \beta F(\lambda )=0.
\label{Fun}\end{equation}
Consider the function $F=F(\xi)$ defined in (\ref{Fun}). It is discontinuous at
the points $ \xi =\lambda _{n}$ and is a monotonically growing function
elsewhere. Note that the summation involves one negative eigenvalue $\lambda
_{1}$, the next (smallest) eigenvalue $\lambda _{3}$ being positive. Then,
since the generalized eigenvalue $\lambda $ defined in Eq.~(\ref{geneig}) is a
zero of the function $F$, we have $\mathrm{sgn}\lambda =-\mathrm{sgn}\{F(0)\}$.
Therefore, we must demand $F(0)<0$ for the soliton stability. Note that
\begin{equation}
F(0)=\langle (U,V)|\Lambda _{1}^{-1}|\left(
\begin{array}{c}
U \\
V
\end{array}
\right) \rangle \,.  \label{F0}
\end{equation}
On the other hand, the differentiation of Eqs. (\ref{E4U})-(\ref{E4V}) with respect
to $\mu $ yields (note that the operator $\Lambda_1$ is even with respect to the
substitution $x\to -x$ )
\begin{equation}
\Lambda _{1}\left(
\begin{array}{c}
{\partial U}/{\partial \mu } \\
{\partial V}/{\partial \mu }
\end{array}
\right) =\left(
\begin{array}{c}
U \\
V
\end{array}
\right) ,  \label{dUVdmu}
\end{equation}
or
\[
\Lambda _{1}^{-1}\left(
\begin{array}{c}
U \\
V
\end{array}
\right) =\left(
\begin{array}{c}
{\partial U}/{\partial \mu } \\
{\partial V}/{\partial \mu }
\end{array} \right).
\]
Finally, substitution of the latter result into Eq. (\ref{F0}) leads to the VK
criterion for the soliton stability:
\begin{equation}
F(0)=\langle (U,V)|\Lambda _{1}^{-1}|\left(
\begin{array}{c}
U \\
V
\end{array}
\right) \rangle =\langle (U,V)|\left(
\begin{array}{c}
{\partial U}/{\partial \mu } \\
{\partial V}/{\partial \mu }
\end{array}
\right) \rangle =\frac{1}{2}\frac{\partial N}{\partial \mu }<0.  \label{crit}
\end{equation}

It should be pointed out that, in the course of the derivation of the stability
criterion (\ref{crit}) for the in-phase soliton solutions of
Eqs.~(\ref{E4U})-(\ref{E4V}), we used the following two facts about the
operator $\Lambda _{1}$: ($i$) there is only one (non-degenerate) negative
eigenvalue, and ($ii$) the corresponding eigenfunction is not orthogonal to the
soliton solution $(U,V)$ (otherwise the lowest generalized eigenvalue $\lambda$
coincides with the negative eigenvalue of $\Lambda _{1}$). With these two
conditions satisfied, one can use the VK stability criterion for other in-phase
soliton solutions of Eqs. (\ref{E1})-(\ref{E2}). Below, this fact is used for
the stability analysis of the sech-type solitons (\ref {exactinph}), and the
solutions bifurcating from the sech-type solitons at $\mu =\mu
_{\mathrm{bif}}$.

\medskip \noindent \textsl{II. Stability analysis of the in-phase solitons for
$a<0$}

First of all, let us consider the special case when $\mu _{0}$ is not an
independent parameter, but is the function of $a$ considered above, i.e. $\mu
_{0}=\sqrt{-a}-1/\sqrt{-a} = \mu _{0}(a)$, see Eq. (\ref{max1}). To derive the
expression for the bifurcation point (\ref{bif}), we note that, in general, at
such a point two branches of the soliton solutions $(U^{(1)},V^{(1)})$ and
$(U^{(2)},V^{(2)})$ coincide, hence due to Eq. (\ref{dUVdmu}) we have
\begin{equation}
\Lambda _{1}\left[ \left(
\begin{array}{c}
{\partial U^{(1)}}/{\partial \mu } \\
{\partial V^{(1)}}/{\partial \mu }
\end{array}
\right) -\left(
\begin{array}{c}
{\partial U^{(2)}}/{\partial \mu } \\
{\partial V^{(2)}}/{\partial \mu }
\end{array}
\right) \right] =0.  \label{zeromode}
\end{equation}
Therefore, the bifurcation point is characterized by the appearance of the
second (nodeless) zero mode of the operator $\Lambda _{1}$, in addition to the
zero mode given by Eq. (\ref{zeroL1}). Such a zero mode $(X_{1},X_{2})^{T}$ can
be easily found analytically for the sech-type solitons (\ref{exactinph}); in
this case, it is a solution to the system
\begin{eqnarray*}
&&\frac{\mathrm{d}^{2}X_{1}}{\mathrm{d}x^{2}}+\left[ \mu
+3A_{1}^{2}\mathrm{sech}^{2}\left( \frac{A_{1}x}{\sqrt{2}}\right) \right] X_{1}+X_{2}=0, \\
&&\frac{\mathrm{d}^{2}X_{2}}{\mathrm{d}x^{2}}+\left[ \mu -\mu _{0}+3A_{1}^{2}
\mathrm{sech}^{2}\left( \frac{A_{1}x}{\sqrt{2}}\right) \right] X_{2}+X_{1}=0.
\end{eqnarray*}
Solving these equations, we obtain the expression (\ref{bif}) for the
bifurcation point $\mu _{\mathrm{bif}}$, and the corresponding zero mode
\begin{equation}
\left(
\begin{array}{c}
X_{1} \\
X_{2}
\end{array}
\right) =\left(
\begin{array}{c}
1 \\
-\sqrt{-a}
\end{array}
\right) \mathrm{sech}^{2}\left( \frac{A_{1}x}{\sqrt{2}}\right) ,\quad
A_{1}=\left( \frac{2(1-a)}{3\sqrt{-a}}\right) ^{1/2}.  \label{2zero}
\end{equation}

Now, we aim to show that the sech-type solitons (\ref{exactinph}) are stable in
the interval
\begin{equation}
\mu _{\mathrm{bif}}<\mu <\mu _{\max }^{(1)},
\label{stable}
\end{equation}
where $\mu _{\max }^{(1)}=-1/\sqrt{-a}$ [see Eq. (\ref {max1})], and unstable
otherwise [we remind that here $\mu _{0}=\mu _{0}(a)$]. We need to determine
the direction of the shift of the zero eigenvalue for $\mu \neq \mu
_{\mathrm{bif}}$ as we move along a given branch of the soliton solutions. To
this end, we can use the perturbation theory for eigenvalues of linear
operators. A shift of the chemical potential from the bifurcation point by a
small amount $\epsilon $, $\mu = \mu _{\mathrm{ bif}}+\epsilon $, results in a
perturbation of the eigenfunction (\ref{2zero}) and the corresponding
eigenvalue $\tilde\lambda =0+\epsilon l_{1}+\mathcal{O} (\epsilon ^{2})$.
Substitution of the perturbed eigenfunction,
\[
\left(
\begin{array}{c}
\tilde{X}_{1} \\
\tilde{X}_{2}
\end{array}
\right) =\left(
\begin{array}{c}
X_{1} \\
X_{2}
\end{array}
\right) +\epsilon \left(
\begin{array}{c}
\chi _{1} \\
\chi _{2}
\end{array}
\right) +\mathcal{O}(\epsilon ^{2}),
\]
into the eigenvalue problem
\begin{equation}
\Lambda _{1}\left(
\begin{array}{c}
\tilde{X}_{1} \\
\tilde{X}_{2}
\end{array}
\right) =\tilde\lambda \left(
\begin{array}{c}
\tilde{X}_{1} \\
\tilde{X}_{2}
\end{array}
\right) \,,
\label{perturbeig}\end{equation}
and keeping only the first-order terms in $\epsilon $, we derive the
following equation for $l_{1}$,
\begin{equation}
\Lambda _{1}\left(
\begin{array}{c}
\chi _{1} \\
\chi _{2}
\end{array}
\right) =l_{1}\left(
\begin{array}{c}
X_{1} \\
X_{2}
\end{array}
\right) +\left(
\begin{array}{cc}
1+6U(\partial U/\partial \mu ) & 0 \\
0 & 1-6aV(\partial V/\partial \mu )
\end{array}
\right) \left(
\begin{array}{c}
X_{1} \\
X_{2}
\end{array}
\right) ,  \label{lam1}
\end{equation}
where the soliton solutions $U$ and $V$ are given by Eq. (\ref{exactinph}), and
the operator $\Lambda _{1}$, together with $U$, $V$ and their derivatives, are
taken at $\mu =\mu _{\mathrm{bif}}$. The left multiplication of Eq.
(\ref{lam1}) by $(X_{1},X_{2})$ and integration over $x$ leads to the following
expression for the sign of $l_{1}$:
\begin{equation}
\mathrm{sgn}\,l_{1}=-\mathrm{sgn}\int \mathrm{d}x\,\left\{ \left( 1+6U\frac{
\partial U}{\partial \mu }\right) X_{1}^{2}+\left( 1-6aV\frac{\partial V}
{\partial \mu }\right) X_{2}^{2}\right\} .  \label{sgnlam1}
\end{equation}
The expressions (\ref{lam1}) and (\ref{sgnlam1}) are valid for the two branches
of the soliton solutions which collide at the bifurcation point $\mu
_{\mathrm{bif}}$ (\ref{bif}).

We now apply Eq. (\ref{sgnlam1}) to the sech-type solitons of
Eq.~(\ref{exactinph}). We have
\[
U=A_{1}\mathrm{sech}\,y,\quad \frac{\partial U}{\partial \mu }=-2\mathrm{sech}
^{2}y(1-y\mathrm{tanh}y),\quad V=\frac{U}{\sqrt{-a}},\quad y\equiv \frac{A_{1}x
}{\sqrt{2}},
\]
thus
\[
\mathrm{sgn}\,l_{1}=-\mathrm{sgn}\int
\mathrm{d}y\,(1-6\mathrm{sech}^{2}y[1-y\mathrm{ tanh}y])\mathrm{sech}^{4}y>0.
\]
Therefore, noticing that $\mathrm{sgn}\tilde\lambda =\mathrm{sgn}(\epsilon
l_{1})$, we arrive at the following formula for the sign of the eigenvalue as
we move along the sech-type branch of the soliton solutions:
\begin{equation}
\mathrm{sgn}\tilde\lambda |_{\mathrm{sech}}=\mathrm{sgn}(\mu -\mu
_{\mathrm{bif}}).
\end{equation}
Thus, we have $\tilde\lambda >0$ for $\mu >\mu _{\mathrm{bif}}$. As the result,
the nodeless eigenfunction from Eq.~(\ref{perturbeig})  with the property
$\tilde{X}_{1}\tilde{X}_{2}<0$ corresponds to a positive eigenvalue, and
$\Lambda _{1}$ has only one non-degenerate negative eigenvalue for $\mu
_{\mathrm{bif}}<\mu <\mu _{\max }^{(1)}$.

To complete the proof of the stability of the sech-type solitons in the
interval (\ref{stable}), we note that the eigenfunction corresponding to the
(only) negative eigenvalue $\lambda _{1}$ of $\Lambda _{1}$ cannot be
orthogonal to the soliton solution $(U,V)$ (\ref{exactinph}). Indeed, such an
eigenfunction $(Z_{1},Z_{2})^T$ has no nodes and satisfies the condition
$Z_{1}Z_{2}>0$ at $\mu =\mu _{\mathrm{bif}}$, as it is orthogonal to the zero
mode (\ref{2zero}). Thus, to become orthogonal to the soliton solution $(U,V)
$, one of the components of this eigenfunction should first pass through zero
at some value of $\mu $ which is impossible.

For $\mu _{0}=\mu _{0}(a)$, the stability properties of the in-phase solitons
bifurcating from the sech-type soliton solutions at $\mu =\mu _{ \mathrm{bif}}$
are affected by the turning-point bifurcation mentioned in
section~\ref{numerics} (see, for instance, the top panel of figure~7). To find
where the VK criterion applies to these solitons, we note that, at the
collision bifurcation point $\mu_\mathrm{bif}$, these soliton solutions share
the operator $\Lambda _{1}$ with the sech-type solitons and at the turning
point one of the (two) negative eigenvalues of $\Lambda _{1}$ passes through
zero and becomes positive, as we go from the upper part of the branch to the
lower one [this is due to the fact that the function $F(\xi )$ makes a jump
from $-\infty $ to $+\infty $ at $\xi =0$ as we pass the turning point
downwards, due to $\partial N/\partial \mu=2F(0)$]. Therefore, as we move along
this (i.e., bifurcating) branch from the bifurcation point $\mu
_{\mathrm{bif}}$, the zero eigenvalue corresponding to the nodeless zero mode
(\ref{2zero}) of $\Lambda _{1}$ becomes negative, but at the turning point it
passes through zero and becomes positive. Therefore, the VK criterion applies
to the lower part of the branch, while the solitons of the upper part are
always unstable (there is one negative generalized eigenvalue for them).

To complete the consideration of the special case with $\mu _{0}=\mu _{0}(a)$,
we note that, at the bifurcation point $\mu =\mu _{ \mathrm{bif}}$, the numbers
of atoms for the two branches of the in-phase solitons vs. chemical potential,
say $N_1=N_{1}(\mu )$ and $N_2=N_{2}(\mu )$, have equal slopes: $\partial
N_{1}/\partial \mu =\partial N_{2}/\partial \mu $, which is a straightforward
corollary of Eq. (\ref{dUVdmu}). Indeed, the zero mode (\ref{2zero}) of the
operator $\Lambda _{1}$, also defined as
\[
\left(
\begin{array}{c}
X _{1} \\
X _{2}
\end{array}
\right) \equiv \left(
\begin{array}{c}
{\partial U^{(1)}}/{\partial \mu } \\
{\partial V^{(1)}}/{\partial \mu }
\end{array}
\right) -\left(
\begin{array}{c}
{\partial U^{(2)}}/{\partial \mu } \\
{\partial V^{(2)}}/{\partial \mu }
\end{array}
\right),
\]
is orthogonal to the soliton solution $(U,V)^T$ taken at the bifurcation point
$\mu _{\mathrm{bif}}$, as it is the right-hand side of Eq. (\ref{dUVdmu}).
Thus,
\[
\frac{\partial N_{1}}{\partial \mu }-\frac{\partial N_{2}}{\partial \mu } =\int
\mathrm{d}x\,\left( U X_{1} + V X _{2}\right) =0.
\]
Geometrically, it means that the in-phase solitons undergo a \textit{
tangential} (one-sided cusp-like) bifurcation. This fact was already
illustrated numerically in the previous section.

The stability properties of the soliton solutions for $\mu _{0}\neq \mu
_{0}(a)$ can be summarized as follows. According to their presentation in
Figs.~7, 8 and 9(c)-(d), we will refer to the two branches of the soliton
solutions as the thin and the thick branch, respectively, where the latter one
corresponds to the sech-type solitons (\ref{exactinph}) for $\mu _{0}=\mu
_{0}(a)$.

First of all, the VK criterion applies to the solitons belonging to the thick
branch for all values of $\mu $, thus they are stable when the number of atoms
decreases with increase of the chemical potential. Second, to understand the
stability of the solitons belonging to the thin branch, we note that, at the
turning point, one of the negative eigenvalues of $\Lambda _{1}$ passes through
zero as one goes from the upper part of the branch to the lower part, similar
as in the special case $\mu _{0}=\mu _{0}(a)$. Thus, the solitons belonging to
the upper branch are unstable, as the operator $\Lambda _{1}$ has two negative
eigenvalues. The VK criterion applies to the lower branch if the (only)
negative eigenvalue of $\Lambda _{1}$ there has the eigenfunction  which is
non-orthogonal to the soliton solution proper. It is enough to verify this
condition numerically just at one point. The outcome is that the VK criterion
is indeed applicable to the lower part of the thin branch.

The analytical results on the soliton stability  were checked by direct
numerical solution of the linear eigenvalue problem (\ref{eigenprob}), and by
counting the negative eigenvalues of the operator $\Lambda _{1}$ (we used the
Fourier spectral discretization method with up to 256 grid points in the LAPACK
routines of Matlab). As the outcome, it has been verified that the analytical
results completely agree with numerical ones in all the cases.

\subsection{Evolution of perturbed unstable solitons}

Here we report results of direct numerical simulations of Eqs.
(\ref{E1})-(\ref{E2}) with perturbed soliton solutions as the initial
conditions. We have used the Fourier spectral discretization method in the $x$
coordinate combined with the leap-frog time-stepping scheme. The stability of
the scheme itself was guaranteed by selecting a small enough time step (we had
$\Delta t=0.001$), such that the stability domain of the leap-frog
time-stepping method contains the eigenvalues of the spacial discretization
operator multiplied by $\Delta t$. The radiation was absorbed by introducing a
smooth distributed damping (smooth damping is necessary for stability of a
numerical scheme based on the spectral methods), which had a negligible effect
on the localized solution.

Previously, the results of numerical simulations for the case of negative $a$
in the context of the  nonlinear dual-fiber optical model were reported in
Ref.~\cite{A3}, where $a=-1$ and $\mu _{0}=0$ were used. Here, our main
interest is the evolution of unstable solitons in the most interesting case,
when $a\geq 0$, corresponding to nonlinearities of the opposite signs in the
two condensates.

First of all, we note that the in-phase solitons may have only one unstable
mode, i.e., the linear eigenvalue problem (\ref{eigenprob}) can have only one
negative eigenvalue $\Omega ^{2}$ (or equivalently, one imaginary
eigenfrequency $\Omega $). For $a=-1$ and $\mu _{0}=0$, similar result was
reported in Ref.~\cite{A2}.

In the context of the stability analysis presented above, it is easy to
understand why there is just one unstable mode. Indeed, where the VK criterion
applies, and the solitons become unstable due to the positive slope,
$\mathrm{d}N/\mathrm{d}\mu >0$, this means that one generalized eigenvalue of
the problem (\ref{geneig}) has passed zero and become negative with the change
of sign of $F(0)$, see Eq. (\ref{crit}). For the soliton solutions
corresponding to the top part of the thin branches in Figs.~7-9  the VK
criterion does not apply, but note, however,  that the slope is negative. Thus,
the only possible unstable mode is due to the appearance of the second negative
eigenvalue in the spectrum of $\Lambda _{1} $ and, hence, a single negative
generalized eigenvalue of the linear problem (\ref{geneig}). The fate of the
corresponding (second) negative eigenvalue of $\Lambda_1$ in this case depends
on whether there is a collision bifurcation or not. In the former case, the
negative eigenvalue must go to zero as we approach the bifurcation point $\mu
_{\mathrm{bif}}$, see Fig.~7, since at $\mu =\mu _{\mathrm{bif}}$ this
eigenvalue is zero (the stability criterion applies to the branch of the
sech-type solitons on one side from the bifurcation point).  In the latter
case, when there is no collision bifurcation, the negative eigenvalue tends to
$-\infty $ as we move along the top part of the thin branch away from the
turning point, see for instance, the top part of figure~8. In both cases, this
negative eigenvalue also goes to zero as we approach the turning point (i.e.,
as we move in the opposite direction).

The fact that there is just one unstable mode, i.e., one unstable direction for
the growth of a small perturbation, simplifies the task of understanding the
fate of the unstable soliton solutions subject to perturbations: it is
sufficient to solve the initial-value problem for Eqs.~(\ref{E1})-(\ref{E2}),
taking as the initial condition  the soliton plus a perturbation proportional
to its unstable mode.

Here we note that though the perturbation based on the unstable mode is complex
(consult appendix~\ref{the mode} for more details), it is sufficient to use
just its real part. This is due to the fact that the soliton solutions are
given by the real functions and, hence, only the real part of the perturbation,
e.g., proportional to $(X_u,X_v)^T$ from equation (\ref{eigenprob}), affects
the number of atoms in the first order approximation. Indeed, for a
perturbation $u - U = \epsilon(X_u+iY_u)$ and $v - V = \epsilon(X_v+iY_v)$,
where $U$ and $V$ are the soliton components, the corresponding variation of
the number of atoms reads
\begin{equation}
\Delta N = 2\epsilon\int\limits_{-\infty}^\infty\mathrm{d}x\,
\left(UX_u +VX_v\right) + \mathcal{O}(\epsilon^2).
\end{equation}
Below by the small perturbation (or the unstable mode) we mean the real part.

We now focus on the case of $a\geq 0$. We found that the evolution of an
unstable soliton is determined by the shape of the stability curve $N=N(\mu)$
and the sign of the small perturbation in the form of the unstable mode. There
are two distinct cases, which correspond to the shapes shown in Figs.~2(c) and
the left panel of Fig. 5, respectively.

We first consider the evolution of the unstable solitons in the former case.
Figure~10 shows the (unstable) soliton and its unstable mode for this case. We
chose $a=0$ (no nonlinearity in the $v$-subsystem) and $\mu_0=-1$, but the
evolution is similar for other values of these parameters, provided that they
give rise to a similar shape of the curve $N=N(\mu )$.  For example, it is so
for $a$ and $\mu_0$ of Fig.~4  (and in any other case of $a>0$, when the
$v$-condensate has atoms with repulsion, but the shape of $N(\mu )$ is similar
to the one given by Fig.~2(c)).

\begin{figure}[f]
\psfig{file=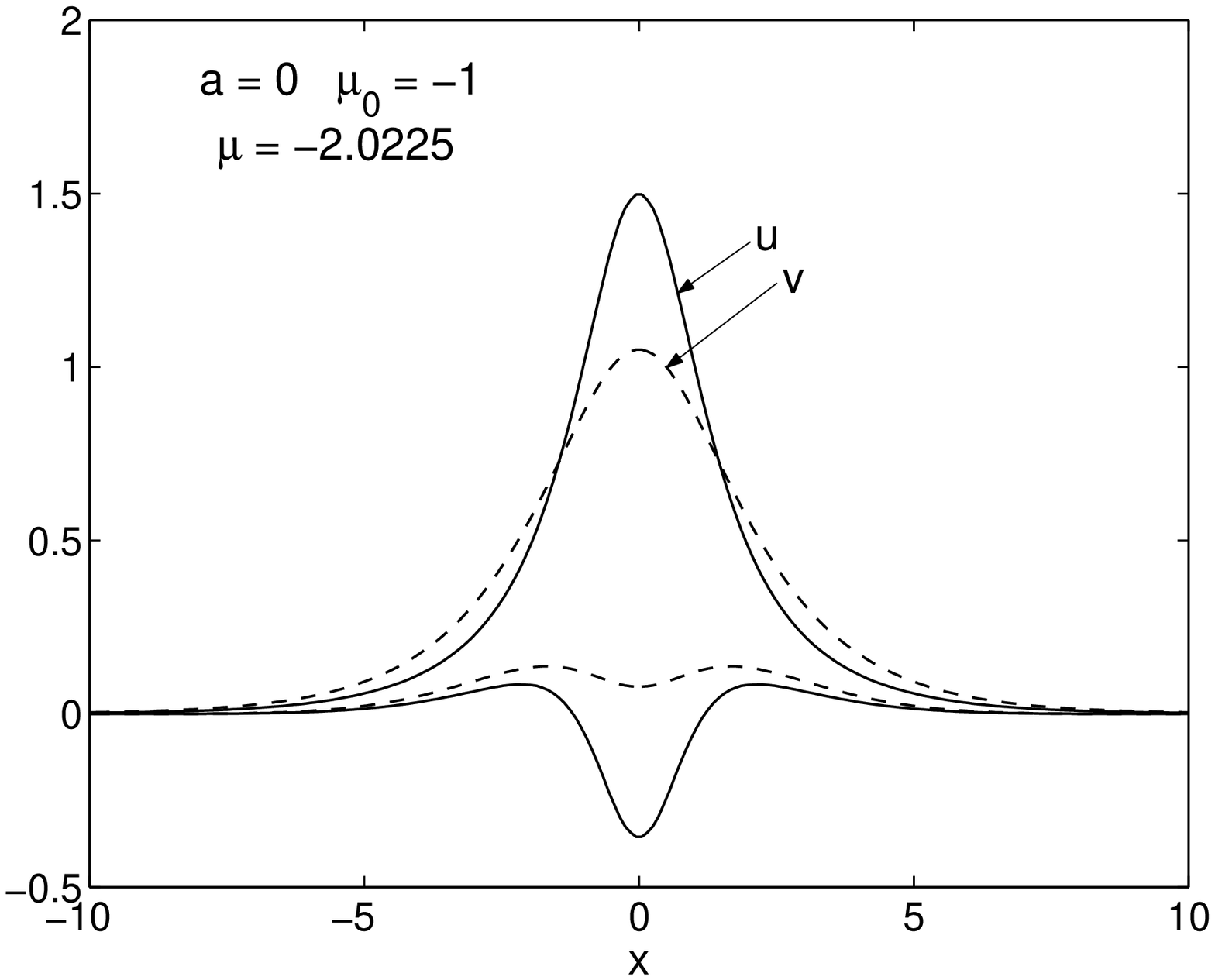,height=0.45\textwidth,width=0.45\textwidth}
 \caption{The unstable soliton and the real part of the single eigenmode of small
perturbations which gives rise to instability.}
\end{figure}

Figure~11 illustrates the evolution of the unstable soliton solution, where the
top picture indicates the direction of the evolution if one adds (the case
denoted by 1) or subtracts (2) the unstable mode, as it was specified above.
For instance, in scenario~1 there were, initially, more atoms trapped in the
$u$-condensate, but this distribution of atoms in the condensates is unstable.
For such initial conditions, the corresponding attracting configuration has
more atoms in the $v$-condensate, as the arrow~1 indicates in the top part of
Fig.~11.

\begin{figure}[f]
\psfig{file=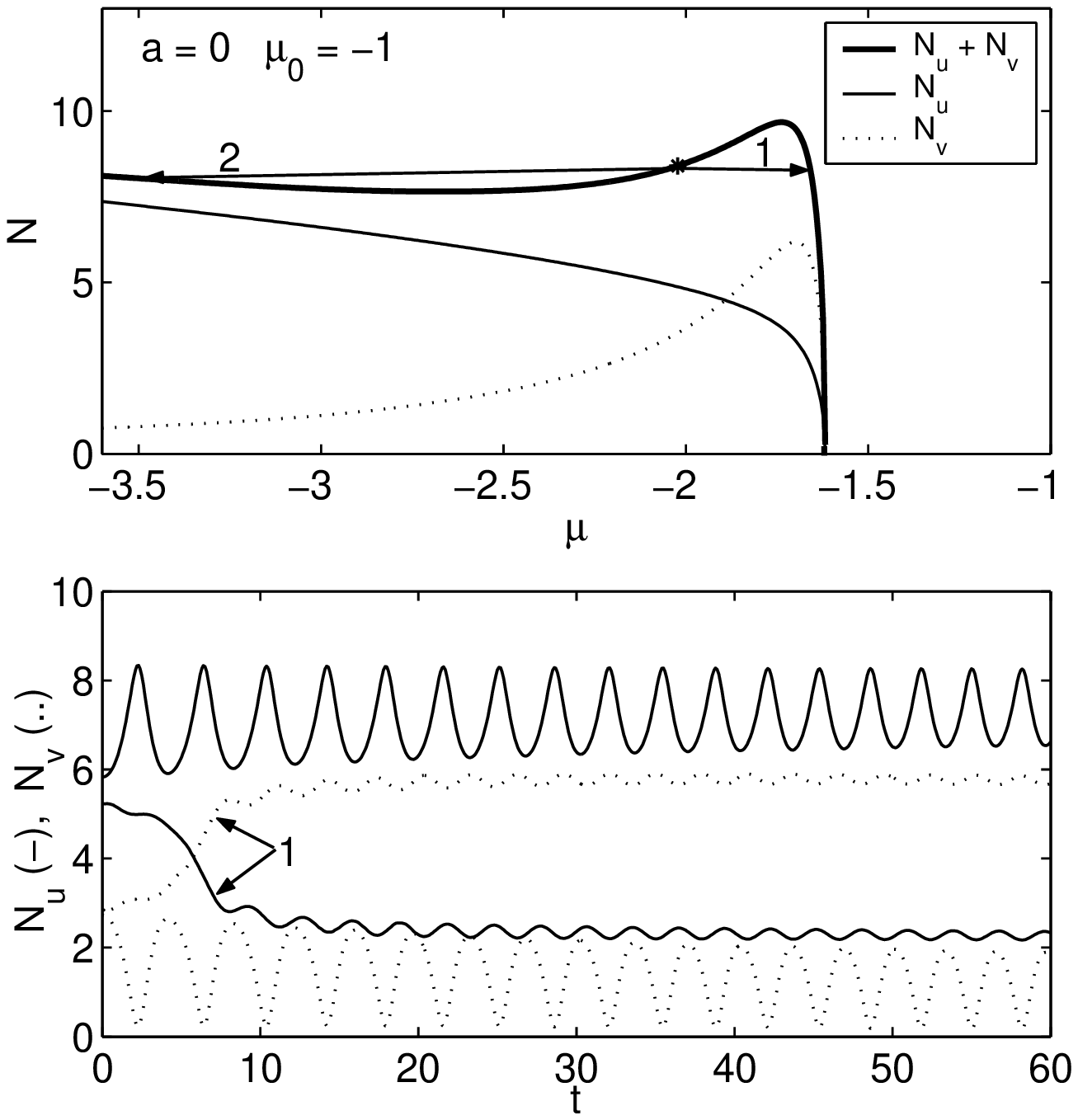,height=0.6\textwidth,width=0.6\textwidth} \caption{ Evolution
of the unstable perturbed soliton. In the top panel, the general direction of the
evolution is indicated relative to the curve $N=N(\mu)$. The cases 1 and 2
correspond, respectively, to the addition and subtraction of the unstable mode (the
one shown in Fig. 10), i.e., to the small perturbation which equals the unstable
eigenmode multiplied by a small positive or negative amplitude. The bottom panel:
the corresponding evolution of the numbers of particles in the two condensates. }
\end{figure}

The picture in the bottom part of Fig.~11 shows the evolution of
the two numbers of atoms, initiated by the instability of the
original soliton. As it follows from this picture, in both cases
the instability gives rise to a breather-like state with
persistent internal vibrations. However, the amplitude of the
vibrations is dependent on the slope of the stability curve: for
the steeper slope the  amplitude is smaller.

Now let us consider what happens to the unstable soliton when the stability
curve $N=N(\mu )$ has the shape of that shown in the left panel of Fig.~5. In
this case, a scenario of the type 2 in terms of Fig. 11 (with formation of a
breather) is also observed. However, a scenario of the type 1 does not take
place in this case. It is replaced by complete decay of the soliton into
radiation, as is shown in Fig.~12, which pertains to the case $a=0.25$ and $\mu
_{0}=-1$, the curve $N(\mu )$ being in this case similar to the one in the left
panel of Fig.~5.

\begin{figure}[f]
\psfig{file=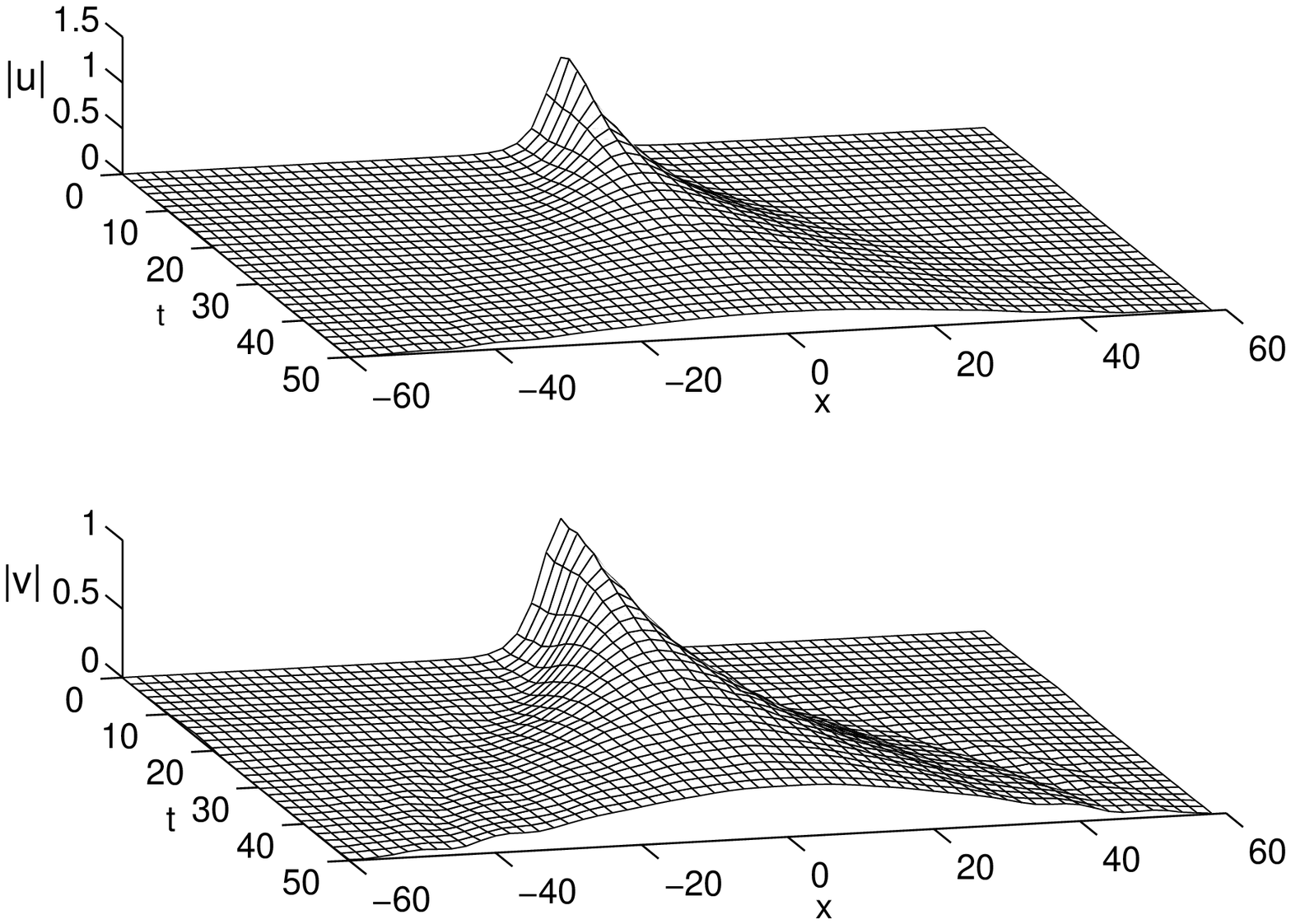,height=0.6\textwidth,width=0.6\textwidth}
\caption{The decay of an unstable soliton into radiation in the
case when $\mathrm{d}N/\mathrm{d}\protect\mu \to\infty$ as
$\protect\mu\to\protect\mu_{ \mathrm{max}}$. This evolution
corresponds, for instance, to the stability curve shown in the
left panel of Fig.~5.}
\end{figure}

\section{Conclusion}

We have found novel soliton states for the system of two linearly coupled NLS
equations, which describe two Bose-Einstein condensates trapped in an
asymmetric double-well potential, in particular, when the atomic interactions
have opposite signs in the two condensates. This system realizes an interplay
between dispersion (corresponding to the kinetic energy in the condensates),
attractive and repulsive nonlinearities, linear coupling, and the
chemical-potential difference between the two traps.

We have found stable soliton solutions with  almost all atoms ($\simeq 96\%$)
being trapped in the condensate with weak repulsive interaction, which is a
novel self-trapping phenomenon in a system of two weakly coupled condensates
having the scattering lengths of the opposite signs. An unusual bifurcation was
found too, when a soliton merges with the zero background, having its amplitude
vanishing and width diverging, while the number of atoms trapped in the soliton
remains finite.

The stability of solitons was studied in detail, and evolution of the unstable ones
was investigated by means of direct numerical simulations. The outcome is that the
unstable solitons either  give rise to breathers or completely decay into
incoherent waves (radiation). Our results for the two-component solitons are also
relevant to the study of coherent atomic tunnelling (see, for instance,
Refs.\cite{tunnel1,tunnel2}), which has attracted a lot of attention in the context
of the Bose-Einstein condensation.

New results concerning the two-component solitons and their stability were also
obtained for the asymmetric model in which the self-interaction is attractive
in both components. This model has direct applications to the dual-core
nonlinear optical fibers. Although it was studied in many works, new results
for this model have been obtained here.

The investigation of the solitons in linearly coupled quasi-one-dimensional
condensates in a double-well potential can be continued in several directions.
These include the study of breathers, and collisions between moving solitons. In
particular, boundary effects will have to be taken into account for moving
solitons, i.e., the form of the longitudinal potential will enter the stage.
Moreover, the antisymmetric (out-of-phase) and higher-order (multi-humped)
solitons, although unstable, may play an important role in the quantum tunnelling
phenomena, since the solution may oscillate about these solitons during a finite
time. These aspects of the soliton dynamics will be addressed elsewhere.

\section{Acknowledgements}
The research by V.S. was supported by the Funda\c{c}\~{a}o de Amparo \`{a}
Pesquisa do Estado de S\~{a}o Paulo (FAPESP), Brazil. B.A.M. appreciates
hospitality of Instituto de Fisica Te\'{o}rica at Universidade Estadual
Paulista (S\~{a}o Paulo, Brazil).

\appendix

\section{Derivation of the coupled-mode equations}
\label{deriv}

Let us outline the derivation of the coupled-mode equations (\ref{E1})-(\ref{E2}).
We will use the  tilde for  the physical variables in the GP equation  to
distinguish them from the corresponding dimensionless variables of the system
(\ref{E1})-(\ref{E2}). The crucial observation is that the modulus of the order
parameter (single-atom wave function) is exponentially small in the barrier region,
see figure~1 of section~\ref{intro}. This allows us to approximate the solution to
the corresponding GP equation with the double-well potential
$V_{\mathrm{ext}}(\vec{\rho})$ strongly confining in the transverse directions
$\vec{\rho}\equiv(y,z)$ (we neglect the longitudinal variation of
$V_{\mathrm{ext}}$),
\begin{equation}
i\hbar \frac{\partial \Psi}{\partial \tilde{t}} =
-\frac{\hbar^2}{2m}\left(\frac{\partial^2 }{\partial \tilde{x}^2}+
\nabla_{\vec{\rho}}^2\right) \Psi +\left(V_{\mathrm{ext}}+g_0|\Psi|^2\right)\Psi,
\label{GP}\end{equation}
as a sum of the factorized wave functions for the two wells:
\begin{equation}
\Psi(\tilde{t},\tilde{x},\vec{\rho}) =
\psi_u(\tilde{t},\tilde{x})\Phi_u(\vec{\rho}) +
\psi_v(\tilde{t},\tilde{x})\Phi_v(\vec{\rho}),
\label{sumgrst}\end{equation}
where $\Phi_u(\vec{\rho})$ and $\Phi_v(\vec{\rho})$ would be the ground-states (for
the transversal degrees of freedom) if the two wells were isolated. Here we point
that $g_0$ is different in the two wells: $g^{(v)}_0/g_0^{(u)}$ has arbitrary value
and  sign. This quotient can be easily managed by technique based on the Feshbach
resonance \cite{fesh} with the magnetic field applied  to only one of the wells.
Inserting the ansatz~(\ref{sumgrst}) in the GP equation~(\ref{GP}) and using the
conditions
\begin{equation}
\int\Phi_u\Phi_v\mathrm{d}^2\vec{\rho}\simeq 0,\quad
\int\Phi_u^2\mathrm{d}^2\vec{\rho} = \int\Phi_v^2\mathrm{d}^2\vec{\rho}=1,
\label{overlap}\end{equation}
one arrives at the system of two equations for the wave functions describing
the longitudinal evolution of the condensates in the wells:
\begin{equation}
i\hbar \frac{\partial \psi_u}{\partial \tilde{t}} =
-\frac{\hbar^2}{2m}\frac{\partial^2\psi_u }{\partial \tilde{x}^2} + ( \mathcal{E}_u
+ g_u|\psi_u|^2)\psi_u - K\psi_v,
\label{unsc1}\end{equation}
\begin{equation}
i\hbar \frac{\partial \psi_v}{\partial \tilde{t}} =
-\frac{\hbar^2}{2m}\frac{\partial^2\psi_v }{\partial \tilde{x}^2} + (\mathcal{E}_v
+ g_v|\psi_v|^2)\psi_v - K\psi_u.
\label{unsc2}\end{equation}
Here the parameters are defined as follows (by changing the sign of either $\Phi_u$
or $\Phi_v$, if necessary, one can set $K>0$)
\[
\mathcal{E}_{u,v} =\int\left(\frac{\hbar^2}{2m}(\nabla_{\vec{\rho}}\Phi_{u,v})^2
+V_{\mathrm{ext}}\Phi_{u,v}^2\right)\mathrm{d}^2\vec{\rho},\quad g_{u,v} = \int
g_0\Phi_{u,v}^4\mathrm{d}^2\vec{\rho},
\]
\begin{equation}
K=-\int\left(\frac{\hbar^2}{2m}(\nabla_{\vec{\rho}}\Phi_{u})(\nabla_{\vec{\rho}}\Phi_v)
 + \Phi_u V_{\mathrm{ext}}\Phi_{v}\right)\mathrm{d}^2\vec{\rho}.
\label{Aparam}\end{equation}
(Note that though the overlap of the ground states $\Phi_u$ and $\Phi_v$ as in
Eq.~(\ref{overlap}) is negligible, the coupling coefficient $K$ is \emph{not}, due
to the local maximum of the potential $V_{\mathrm{ext}}(\vec{\rho})$ in the overlap
region, see Fig.~1 of section~\ref{intro}). Finally, assuming that $g_u<0$ (i.e.,
attractive atomic interaction in the $u$-condensate) and setting
\begin{equation}
\psi_{u} = \sqrt{\frac{K}{-g_u}}\exp\left(-\frac{i\mathcal{E}_u
t}{\hbar}\right)u,\quad \psi_{v} =
\sqrt{\frac{K}{-g_u}}\exp\left(-\frac{i\mathcal{E}_u
t}{\hbar}\right)v,
\label{psitouv}\end{equation}
in Eqs.~(\ref{unsc1})-(\ref{unsc2}) one arrives at the dimensionless system
(\ref{E1})-(\ref{E2}) with $a = -g_v/g_u$ and $\mu_0 =
(\mathcal{E}_v-\mathcal{E}_u)/K$, where the space and time variables of the system
(\ref{E1})-(\ref{E2}) are given as follows $x = (\sqrt{2mK}/\hbar)\tilde{x}$ and $t
= (K/\hbar) \tilde{t}$.

Recalling the well-known expression for the coupling coefficient $g^{(u)}_0 =
4\pi\hbar^2 a^{(u)}_s/m$, where $a^{(u)}_s$ is the scattering length in the
$u$-condensate, and using the definition of $g_u$ (\ref{Aparam}) in the relation
(\ref{psitouv}) we conclude that the actual (i.e., physical) numbers of atoms are
related to the numbers of particles in the model (\ref{E1})-(\ref{E2}) as follows
\begin{equation}
\widetilde{N}_{u} =
\frac{\left[\int\Phi^4_u\mathrm{d}^2\vec{\rho}\,\right]^{-1}}{8\pi
d_0|a^{(u)}_s|}\int\mathrm{d}x |u|^2,\quad
 \widetilde{N}_{v} =
\frac{\left[\int\Phi^4_u\mathrm{d}^2\vec{\rho}\,\right]^{-1}}{8\pi
d_0|a^{(u)}_s|}\int\mathrm{d}x |v|^2,
\label{Nphys}\end{equation}
where we have introduced the characteristic overlap distance by $d_0 \equiv
\hbar/\sqrt{2mK}$.

In the above derivation it was assumed that the transverse profile of the order
parameter is determined by the linear part of the r.h.s. in equation (\ref{GP}).
Such approximation is valid for the weak atomic interaction:
\begin{equation}
\frac{\mathrm{max}(|g^{(u)}_0|,|g^{(v)}_0|)\widetilde{N}}{d^2_\perp d_\parallel}
\ll \frac{\hbar \omega_\perp}{2},
\label{condN}\end{equation}
where $d_\perp = (\hbar/m\omega_\perp)^{1/2}$ and $d_\parallel$ are the transverse
and longitudinal sizes of the condensate, with the transverse size being given by
the oscillator length, $\widetilde{N}$ is the (physical) total number of atoms, and
$\omega_\perp$ is the external trap frequency (in the transverse dimensions)
without the separation barrier. The r.h.s. of formula (\ref{condN}) is an estimate
on the energy of the linear transverse terms in the GP equation (\ref{GP}).
Substitution of the expression for the coupling coefficients  $g^{(u,v)}_0 =
4\pi\hbar^2 a^{(u,v)}_s/m$ in the condition (\ref{condN}) leads to a more
convenient equivalent form:
\begin{equation}
{8\pi\,\mathrm{max}(|a^{(u)}_s|,|a^{(v)}_s|)}\frac{\widetilde{N}}{d_\parallel} \ll
1.
\label{condN2}\end{equation}
To make an estimate we use the usual values:  $a_s\sim 10^{-9}$m and
$d_\parallel\sim 10^{-4}$m (the latter corresponds to the asymmetry parameter
$\gamma\equiv d_\perp/d_\parallel \sim 0.01$ for $d_\perp\sim 10^{-6}$m, usual for
experiments on the cigar-shaped condensates). Thus $\widetilde{N} \ll 10^{6}$
thereby allowing for up to a hundred thousands of atoms in the condensates.

Due to the scaling $d_\parallel \sim N$, where $N$ is the total number of particles
in the model (\ref{E1})-(\ref{E2}) (compare the widths of the solitons from figs.~3
and 4-5), the condition (\ref{condN2}) is satisfied for all values of the
parameters $\mu_0$ and $a$ in the system (\ref{E1})-(\ref{E2}) and all values of
the chemical potential $\mu$ if it is satisfied just at one point $(a,\mu_0,\mu)$.
Therefore, though the total numbers of atoms do vary by an order of magnitude,
nevertheless, the model system (\ref{E1})-(\ref{E2}), in fact, applies uniformly in
the whole parameter space.

A very rough estimate on the  numbers of trapped atoms can be easily carried out.
Note that the coupling coefficient $g_u$ depends  on the geometry of the trap
through the multiplier
\[
\int \Phi_{u}^4\mathrm{d}^2\vec{\rho}\sim\frac{1}{(d^{(u)}_\perp)^2}.
\]
Here $d^{(u)}_\perp$ is the characteristic transverse size of the $u$-condensate.
To make an estimate on the numbers of trapped atoms we assume: $|a^{(u)}_s|\sim
10^{-9}$m, $d^{(u)}_\perp\sim d_\perp = 10^{-6}$m, and $d_0\sim 0.1d^{(u)}_\perp =
10^{-7}$m, what leads to the transformation coefficients in formula (\ref{Nphys})
of the order $10^3$. Thus this estimate shows that all the stable solitons we
discuss in the present paper correspond to numbers of trapped atoms lying below the
estimated bound of $10^6$.

\section{Unstable mode and the related perturbation of the soliton solution}
\label{the mode}

Here we provide some details on the unstable eigenfunction of the linear
stability problem and the related initial perturbation. Looking for eigenmodes
of the linear system (\ref{linprob}) we write
\begin{equation}
\left(\begin{array}{c}\mathcal{U}_R\\ \mathcal{U}_I\\ \mathcal{V}_R\\
\mathcal{V_I}\end{array}\right) = e^{-i\Omega
t}\left(\begin{array}{c}X_u\\Y_u\\X_v
\\ Y_v\end{array}\right) + c.c.,
\label{B1}\end{equation}
since we need a {\it real} solution. The linear problem (\ref{linprob}) then
reduces to an eigenvalue problem for $\lambda=-i\Omega$, which can be written
as a system of two matrix equations:
\begin{equation}
-i\Omega\left(\begin{array}{c}X_u\\X_v \end{array}\right) = \Lambda_0
\left(\begin{array}{c}Y_u\\Y_v \end{array}\right),\quad
-i\Omega\left(\begin{array}{c}Y_u\\Y_v \end{array}\right) =
-\Lambda_1\left(\begin{array}{c}X_u\\X_v \end{array}\right),
\label{B2}\end{equation}
from which one derives equation (\ref{eigenprob}). Note that $(X_u,X_v)^T$ is
real (as the eigenfunction of the eigenvalue problem (\ref{eigenprob})). Then
from the  second equation in (\ref{B2}) we conclude that the eigenfunction
$(X_u,Y_u,X_v,Y_v)^T$ corresponding to the real eigenvalue (i.e. imaginary
eigenfrequency $\Omega$) is real. Taking this into account and using equations
(\ref{B1})-(\ref{B2}) it is straightforward to deduce that the initial
perturbation  $(u_1,v_1)^T$ of the soliton solution which is based on the
unstable eigenfunction $(X_u,Y_u,X_v,Y_v)^T$ is given as
\begin{equation}
\left(\begin{array}{c}u_1\\v_1 \end{array}\right) = \sigma\left(1 -
\frac{i\Lambda_1}{\mathrm{Im}\Omega}\right)\left(\begin{array}{c}X_u\\X_v
\end{array}\right),
\end{equation}
where $\sigma$ is arbitrary real constant.

\section{Negative eigenvalues of the operator $\Lambda_1$ for $a\ge0$}

\label{App}

To prove that $\Lambda _{1}$ has only one nodeless eigenfunction (and, hence,
only one negative eigenvalue) for $a\ge 0$, we use the fact that the in-phase
soliton solutions correspond to the $k_{1}$-branch of the dispersion relation
(\ref{disp}). The eigenfunction $(X_{1},X_{2})^{T}$ of the operator $\Lambda
_{1}$ corresponding to an eigenvalue $\lambda $ satisfies the following system
of equations,
\begin{eqnarray}
&&\frac{\mathrm{d}^{2}X_{1}}{\mathrm{d}x^{2}}+\left( \mu +\lambda
+3U^{2}\right)
X_{1}+X_{2}=0, \\
&&\frac{\mathrm{d}^{2}X_{2}}{\mathrm{d}x^{2}}+\left( \mu +\lambda -\mu
_{0}-3aV^{2}\right) X_{2}+X_{1}=0,  \label{A2}
\end{eqnarray}
which follow from the definition of $\Lambda _{1}$. We will prove that,
among two possible combinations of nodeless eigenfunctions, $X_{1}X_{2}>0$
and $X_{1}X_{2}<0$, the latter one is not possible for the in-phase solitons
when $a\geq 0$.

The multiplication of Eq. (\ref{A2}) by $\mathrm{d}X_{2}/\mathrm{d}x$ and
integration from $x=0$ to $x=\infty $ give
\begin{equation}
(\mu -\mu _{0}+\lambda )X_{2}^{2}(0)=\int\limits_{0}^{\infty }\mathrm{d}
x\,\left\{
-6aV^{2}X_{2}\frac{\mathrm{d}X_{2}}{\mathrm{d}x}+2X_{1}\frac{\mathrm{d}
X_{2}}{\mathrm{d}x}\right\} .  \label{A3}
\end{equation}
Here we have used that the eigenfunction has no nodes, hence ${\mathrm{d}X_{2}}
/{\mathrm{d}x}=0$ at $x=0$. Taking into account that the condition $k_{1}^{2}>0
$ (see equation (\ref{disp})) implies $\mu -\mu _{0}<0$, and that
$X_{2}(\mathrm{d}X_{2}/\mathrm{d}x)<0$ for $x>0$, we conclude that Eq. (\ref
{A3}) cannot be satisfied for a negative eigenvalue $\lambda $ if $X_{1}(
\mathrm{d}X_{2}/\mathrm{d}x)>0$ for $x>0$, i.e., the operator $\Lambda _{1}$
cannot have an eigenfunction corresponding to a negative eigenvalue and,
simultaneously, having the property $X_{1}X_{2}<0$.


\begin{thebibliography}{99}

\bibitem{exp1}  M. H. Anderson, J. R. Ensher, M. R. Matthews, C. E. Wieman,
and E. A. Cornell, Science \textbf{269}, 198 (1995).

\bibitem{exp2}  C. C. Bradley, C. A. Sackett, J. J. Tollett, and R. G.
Hulet, Phys. Rev. Lett. \textbf{75}, 1687 (1995).

\bibitem{exp3}  M. O. Mewes, M. R. Andrews, N. J. van Druten, D. M. Kurn, D.
S. Durfee, C. G. Townsend, and W. Ketterle, Phys. Rev. Lett. \textbf{77},
416 (1996); \textbf{77}, 988 (1996).

\bibitem{exp4}  M. R. Andrews, C. G. Townsend, H.-J. Miesner, D.S. Durfee,
D. M. Kurn, and W. Ketterle, Science \textbf{275}, 637 (1997).


\bibitem{GPE}  L. P. Pitaevskii, Zh. Eksp. Teor. Fiz. \textbf{40}, 646
(1961) [Sov. Phys. JETP \textbf{13}, 451 (1961); E. P. Gross, Nuovo Cimento
\textbf{20}, 454 (1961); J. Math. Phys. \textbf{4}, 195 (1963). See also a
review: F.~Dalfovo, S. Giorgini, L. P. Pitaevskii, and S. Stringari, Reviews
of Modern Physics, \textbf{71}, 463 (1999).

\bibitem{rev1}  J. R. Anglin and W. Ketterle, Nature \textbf{416}, 211
(2002).


\bibitem{fesh}  S. L. Cornish, N. R. Claussen, J. L. Roberts, E. A. Cornell,
and C. E. Wieman, Phys. Rev. Lett. \textbf{85}, 1795 (2000).


\bibitem{optfiber}  A. Hasegawa and Y. Kodama, \textit{Solitons in Optical
Communications} (Oxford University Press, Oxford, 1995).


\bibitem{dark1}  S. Burger, K. Bongs, S. Dettmer, W. Ertmer, K. Sengstock,
A. Sanpera, G. V. Shlyapnikov, and M. Lewenstein, Phys. Rev. Lett.
\textbf{83}, 5198 (1999).

\bibitem{dark2}  J. Denschlag, J. E. Simsarian, D. L. Feder, C. W. Clark, L.
A. Collins, J. Cubizolles, L. Deng, E. W. Hagley, K. Helmerson, W. P.
Reinhardt, S. L. Rolston, B. I. Schneider, and W. D. Phillips, Science
\textbf{287}, 97 (2000).

\bibitem{dark3}  B. P. Anderson, P. C. Haljan, C. A. Regal, D. L. Feder, L.
A. Collins, C. W. Clark, and E. A. Cornell, Phys. Rev. Lett. \textbf{86},
2926 (2001).

\bibitem{dark4}  S. Burger, L. D. Carr, P. {\"{O}}hberg, K. Sengstock, and
A. Sanpera, Phys. Rev. A \textbf{65}, 043611 (2002).

\bibitem{bright1}  K. E. Strecker, G. B. Partridge, A. G. Truscott, and R.
G. Hulet, Nature \textbf{417}, 150 (2002).


\bibitem{modul}  N. Tsukada, M. Gotoda, Y. Nomura, and T. Isu, Phys. Rev. A
\textbf{59}, 3862 (1999).

\bibitem{tunnellimit}  G. J. Milburn, J. Corney, E. M. Wright, and D. F.
Walls, Phys. Rev. A \textbf{55} 4318 (1997).

\bibitem{tunnel1}  A. Smerzi, S. Fantoni, S. Giovanazzi, and S. R. Shenoy,
Phys. Rev. Lett. \textbf{79}, 4950 (1997).

\bibitem{tunnel2}  S. Raghavan, A. Smerzi, S. Fantoni, and S. R. Shenoy,
Phys. Rev. A \textbf{59}, 620 (1999).


\bibitem{timedep}  F. Kh. Abdullaev and R. A. Kraenkel, Phys. Rev. A \textbf{
62}, 023613 (2001).

\bibitem{modopt}  S. Raghavan and G. P. Agrawal, J. Modern Optics \textbf{47},
1155 (2000).


\bibitem{NewReview}  B.A. Malomed, Progr. Optics \textbf{43}, 69 (2002).

\bibitem{couplmod}  E. A. Ostrovskaya, Yu. S. Kivshar, M. Lisak, B. Hall, F.
Cattani, and D. Anderson, Phys. Rev. A \textbf{61}, 031601 (R) (2000).

\bibitem{nonpert}  R. D'Agosta and C. Presilla, Phys. Rev. A \textbf{65},
043609 (2002).

\bibitem{darkbright}  Th. Busch and J. R. Anglin, Phys. Rev. Lett. \textbf{87},
010401 (2001).


\bibitem{A1}  N. Akhmediev and A. Ankiewicz, Phys. Rev. Lett. \textbf{70},
2395 (1993).

\bibitem{A2}  J. M. Soto-Crespo and N. Akhmediev, Phys. Rev. E \textbf{48},
4710 (1993).

\bibitem{A3}  N. Akhmediev and J. M. Soto-Crespo, Phys. Rev. E \textbf{49},
4519 (1993).

\bibitem{M1}  B. A. Malomed, I. M. Skinner, P. L. Chu, and G. D. Peng, Phys.
Rev. E \textbf{53}, 4084 (1996).

\bibitem{K1}  D. J. Kaup, T. Lakoba, and B. A. Malomed, J. Opt. Soc. Am. B
\textbf{14}, 1199 (1997).

\bibitem{K2}  D. J. Kaup and B. A. Malomed, J. Opt. Soc. Am. B \textbf{15},
2838 (1998).


\bibitem{Forn}  B. Fornberg, \textit{A Practical Guide to Pseudospectral
Methods} (Cambridge University Press, Cambridge, UK, 1996).

\bibitem{Boyd}  J. P. Boyd, \textit{Chebyshev and Fourier Spectral Methods},
Second Edition (DOVER Publications, Inc., New York, 2000).

\bibitem{Tref}  L. N. Trefethen, \textit{Spectral Methods in Matlab} (SIAM,
Philadelphia, PA, 2000).


\bibitem{VK}  M. G. Vakhitov and A. A. Kolokolov, Radiophys. Quant.
Electron. \textbf{16}, 783 (1975).

\bibitem{Zakh}  E. A. Kuznetsov, A. M. Rubenchick, and V. E. Zakharov, Phys.
Rep. \textbf{142}, 103 (1986).


\bibitem{Iooss}  G. Iooss and D. Joseph, \textit{Elementary Stability and
Bifurcation Theory}, 2nd edition, (Springer-Verlag, New York, 1990 ).
\end{thebibliography}
\end{document}